\documentclass[]{aa}
\usepackage{graphicx}
\usepackage{txfonts}
\usepackage{graphics,graphicx,epstopdf,epsfig,url}
\usepackage{relsize}
\usepackage{dcolumn}
\usepackage{bm}
\usepackage[normalem]{ulem}
\usepackage{times}
\usepackage{natbib}
\usepackage{color}
\usepackage{multirow}
\usepackage{hyperref}
\usepackage{braket}
\usepackage[utf8]{inputenc}
\usepackage{xcolor}
\usepackage{hypcap}
\usepackage[autostyle]{csquotes}
\usepackage{textcomp }
\hypersetup{
    colorlinks=true, 
    citecolor=blue, 
    filecolor=black, 
    linkcolor=blue,
    urlcolor= black
}
\usepackage{etoolbox}
\makeatletter
\patchcmd\@combinedblfloats{\box\@outputbox}{\unvbox\@outputbox}{}{%
   \errmessage{\noexpand\@combinedblfloats could not be patched}%
}%
 \makeatother

\def\titletext{The baryon cycle of Seven Dwarfs with superbubble feedback}
\def\codename{\textsc{Gasoline2} }
\def\codenamex{\textsc{Gasoline2}}

\def\shenone{\cite{2014ApJ...792...99S} }
\def\shentwo{\cite{2014ApJ...789L..17M} }

\def\shenonex{\cite{2014ApJ...792...99S}}
\def\shentwox{\cite{2014ApJ...789L..17M}}

\begin{document}
\title{\titletext}

 \author{Mattia Mina\inst{1}
          \and
          Sijing Shen\inst{1}
          \and
          Benjamin Walter Keller\inst{2}
          \and
          Lucio Mayer\inst{3}
          \and
          Piero Madau\inst{4}
          \and
          James Wadsley\inst{5}
          }

\institute{Institute of Theoretical Astrophysics, University of Oslo, 0315 Oslo, Norway\\
\email{mattia.mina@astro.uio.no}\\
\email{sijing.shen@astro.uio.no}
\and
Astronomisches Rechen-Institut, Zentrum für Astronomie der Universit\"at Heidelberg, 69120 Heidelberg, Germany
\and
Center for Theoretical Astrophysics and Cosmology, Institute for Computational Science, University of Zurich, Winterthurerstrasse 190, CH-9057 Zurich, Switzerland
\and
Department of Astronomy and Astrophysics, University of California, 1156 High Street, Santa Cruz, CA 95064, USA
\and
Department of Physics and Astronomy, McMaster University, Hamilton, Ontario, L8S 4M1, Canada
}

\date{Date}

\abstract{
We present results from a high-resolution, cosmological, $\Lambda$CDM simulation of a group of field dwarf galaxies with the ``superbubble'' model for clustered SN feedback, accounting for thermal conduction and cold gas evaporation. The initial conditions and the galaxy formation physics, other than SN feedback, are the same as in \citet{2014ApJ...792...99S}.  
The simulated luminous galaxies have blue colors, low star formation efficiencies and metallicities, and high cold gas content, reproducing the observed scaling relations of dwarfs in the Local Volume. Bursty star formation histories and superbubble-driven outflows lead to the formation of kpc-size DM cores when stellar masses reaches $M_{\rm *} > 10^{6}$ $M_{\odot}$, similar to previous findings. However, the superbubble model appears more effective in destroying DM cusps than the previously adopted ``blastwave'' model, reflecting a higher coupling efficiency of SN energy with the ISM. On larger scale, superbubble-driven outflows have a more moderate impact: galaxies have higher gas content, more extended stellar disks, and a smaller metal-enriched region in the CGM. The two halos with $M_{\rm vir} \sim 10^{9}$ $M_{\odot}$, which formed ultra-faint dwarf galaxies in \citet{2014ApJ...792...99S}, remain dark due to the different impact of metal-enriched galactic winds from two nearby luminous galaxies.
The column density distributions of \ion{H}{i}, \ion{Si}{II}, \ion{C}{IV} and \ion{O}{vi} are in agreement with recent observations of CGM around isolated dwarfs. While \ion{H}{i} is ubiquitous with a covering fraction of unity within the CGM, \ion{Si}{II} and \ion{C}{IV} are less extended.
\ion{O}{VI} is more extended, but its mass is only 11\% of the total CGM oxygen budget, as the diffuse CGM is highly ionised by the UV background. Superbubble feedback produces \ion{C}{IV} and \ion{O}{VI} column densities that are an order of magnitude higher than those with blastwave feedback. Thus, the CGM and DM cores are most sensitive probes of feedback mechanisms.
}


\maketitle

\section{Introduction}
Within the standard $\Lambda$ Cold Dark Matter ($\Lambda$CDM) model, the Universe today consists of photons and ordinary matter ($\sim 5$\%), a cold and collisionless form of matter known as dark matter ($\sim 27$\%) and a cosmological constant $\Lambda$ ($\sim 68$\%) which describes the dark energy component, responsible for the late time accelerated expansion.
The $\Lambda$CDM model has been very successful when describing the large--scale properties of the observed Universe. However, small scales are still a source of unanswered questions.

The first generation of cosmological dark matter-only simulations has shown the existence of severe tensions between theoretical predictions of the $\Lambda$CDM model and astronomical observations.
In the \textit{Core-Cusp} problem, in small and heavily dark matter dominated galaxies, dark matter halos exhibit central density profiles characterised by an almost constant density core \citep{Oh_2015, Agnello_2012, Adams_2014, 10.1111/j.1365-2966.2011.19684.x, Walker_2011, 2008ApJ...681L..13B, 1994ApJ...427L...1F, 1994Natur.370..629M}, which seems to be inconsistent with steeper central densities of dark halos found in dark matter-only simulations of the $\Lambda$CDM model, such as \citet{1991ApJ...378..496D} and \citet{1996ApJ...462..563N}.
Furthermore, several numerical studies, such as \citet{2008MNRAS.391.1685S}, \citet{2009Sci...325..970K}, \citet{2009MNRAS.398L..21S}, \citet{2014MNRAS.438.2578G}, and \citet{2016ApJ...818...10G}, have shown how the $\Lambda$CDM paradigm predicts the formation of a large number of dark matter subhalos, many more than the actual number of observed satellites orbiting M31 and the Milky Way, raising the so-called \textit{Missing Satellite} problem \citep{1999ApJ...524L..19M}.
In addition, the most massive subhalos found in dark matter-only simulations of Milky Way-mass halos have higher peak rotation velocities than the observed Milky Way satellites, and the fact that these seemingly more massive satellites are not observed in the Local Group is often framed as the \textit{Too Big To Fail} problem \citep{2011MNRAS.415L..40B,10.1111/j.1365-2966.2012.20695.x}.

While many alternative dark matter models have been proposed to solve the small-scale crisis \citep[see e.g.,][]{2005PhR...405..279B,2009EPJC...59..557S,2009NJPh...11j5006B}, many recent simulations have shown that environmental effects such as ram pressure and tidal stripping, together with ultraviolet (UV) background heating and stellar feedback, can either regulate gas condensation at the center of galaxies or eject the gas altogether.
Baryonic physics may therefore explain the dwarf low star formation efficiencies, their low abundance compared to the high number of low-mass DM structures predicted by the $\Lambda$CDM model, and provide a solution to the missing satellite problems \citep{2006MNRAS.371..401H, 2008MNRAS.390..920O,2008ApJ...689L..41M,2010Natur.463..203G, 2013MNRAS.432.1989S, 2013ApJ...766...56M,Brooks_Zolotov_2014,2014MNRAS.444..503N,10.1093/mnras/stu2720,2015MNRAS.446.1140T,2016MNRAS.457.1931S,2016ApJ...827L..23W}.
In addition, it was shown that rapid bursts of star formation and outflows due to stellar feedback can dynamically heat the dark matter structure of dwarf galaxies, reducing the central density of dark halos and creating central dark matter cores, providing solutions to the cusp-core problem and related too-big-to-fail problem \citep{2008Sci...319..174M,2010Natur.463..203G,2012MNRAS.421.3464P, 2012ApJ...761...71Z, 2014ApJ...789L..17M, 2015MNRAS.454.2981C,10.1093/mnras/stv2072, 2016MNRAS.459.2573R,2016MNRAS.456.3542T,2017MNRAS.471.3547F}.

These baryonic physical processes have also been shown to be crucial to reproduce realistic dwarf galaxies in the Local Volume. Indeed, modern hydrodynamic simulations of dwarf galaxies \citep[e.g.,][]{Brooks_Zolotov_2014,2014ApJ...792...99S,10.1093/mnras/stv2072,Wheeler2015,2017MNRAS.471.3547F} have achieved great success in reproducing the observed properties, such as stellar mass content \citep[e.g.,][]{2013ApJ...770...57B,2013MNRAS.428.3121M,2014ApJ...784L..14B,2014MNRAS.438.2578G,2017MNRAS.467.2019R,Jethwa2018}, cold gas content, star formation history \citep[e.g.,][]{2009ARA&A..47..371T,Weisz_2011,2014ApJ...789..147W,Brown2014}, mass-metallicity relationship \citep[e.g.,][]{2012AJ....144....4M,2013ApJ...779..102K,2014MNRAS.439.1015K}, stellar and gas kinematics and morphological properties \citep[e.g.,][]{2012AJ....144....4M} in field and satellite dwarf galaxies in the local Universe. With ever increasing resolution, simulations are now capable of probing formation of ultra-faint dwarf galaxies (UFDs) in and making predictions for upcoming deep surveys from telescopes such as the Vera C. Rubin Observatory \citep[e.g.,][Applebaum et al. \it{in prep.}]{2017MNRAS.471.3547F, 2018A&A...616A..96R, Wheeler2019,2019MNRAS.482.1176W, Munshi_2019,Agertz2020}. All these simulations have confirmed that, at least in the regime of classical dwarfs ($M_{\rm vir} \sim 10^{10} M_{\odot}$), galactic outflows driven by stellar feedback is of critical importance to suppress star formation, reduce stellar bulge-to-disk ratio, lower the intestellar medium (ISM) metallicity and enrich the CGM.

Despite the consensus on the importance of galactic winds in dwarf galaxy formation, large uncertainties persist in how the winds are driven. As the ISM physics is typically not resolved in full cosmological simulations \citep[although some simulations are approaching it, see e.g., ][]{Wheeler2019,Agertz2020}, sub-grid feedback models are often used to alleviate the numerical ``overcooling'' problem, whereby energy injected to the ISM is rapidly cooled away due to unresolved ISM structures. Many studies have investigated supernova feedback, perhaps the most important process to drive galactic wind in dwarf galaxies. Different approaches of subgrid models were explored, ranging from kinetic feedback models where wind mass loading and velocities are predetermined \citep{Springel2003,Oppenheimer2006,DallaVecchia2008}, to explicit models where radiative cooling is temporary shut-off \citep{2006MNRAS.373.1074S,Agertz2011, Teyssier2013}, to mechanical feedback where the momentum boost during the unresolved adiabatic phase is calibrated through small-scale ISM simulations and injected into the ISM  \citep{2014MNRAS.445..581H,Kimm_2014,Smith2018}. Simulations of dwarf galaxies in recent years often adopt one of these approaches for supernova feedback. Although many are considered successful in reproducing observations, little work is done to compare these models in full cosmological simulations. It is thus important to gauge how the evolution of simulated dwarf galaxies (especially at the low mass end) is dependent on feedback models, therefore providing more robust model with realistic uncertainties for future observations. 

Moreover, most existing supernova feedback model individual supernova explosions. However, recent studies such as \citet{2013ApJ...777L..12N} and \citet{2014MNRAS.443.3463S}, have pointed out that stars form in clusters and, through multiple repeated supernovae explosions, they generate the so-called superbubbles, which contains hot gas, and the sharp temperature gradients between hot and cold gas make thermal conduction an extremely important process.

Thus, \citet{2014MNRAS.442.3013K} proposed the superbubble model, where the thermal conduction between cold and hot phases of sub-resolution multi-phase particles is included, as well as a stochastic evaporation model for cold gas. As shown by \citet{1988ApJ...324..776M}, thermal conduction describes the mass flow from the cold shell into the hot gas, therefore controls how much gas is heated by feedback. Bubble temperature and mass is determined by the evaporation process, along with radiative cooling, without the need of additional free parameters and naturally avoiding the overcooling problem. In this way, superbubbles efficiently converts feedback energy into thermal and kinetic energy in the hot and the cold phases, respectively. In \citet{2015MNRAS.453.3499K} and \citet{10.1093/mnras/stw2029}, they performed the first cosmological simulations of halos with mass $M_{\rm vir} > 10^{11}~M_{\odot}$ with superbubble feedback, and shown that as this model can launch strong galactic outflows, effectively regulating star formation from early on ($z > 2$). The simulations with supperbubble-driven outflows reproduce many observed properties of $L_{*}$ galaxies. More recently, a new theoretical framework was proposed in \citet{2020MNRAS.493.2149K}, where it was shown that outflows from superbubble feedback have higher entropy and are continuously accelerated in the CGM due to buoyancy. As a consequence, although the winds have higher mass loading and lower velocities, they tend to stay in the CGM in a longer timescale. 

In this work, we extend the investigation to the formation of low-mass dwarf galaxy regime, for halos with $M_{\rm vir} \la 10^{10} M_{\odot}$ by re-simulating the group of seven field dwarf galaxies presented in \citet{2014ApJ...792...99S} with the superbubble feedback model from \citet{2014MNRAS.442.3013K}. The original \citet{2014ApJ...792...99S} work adopted the blastwave feedback \citep{2006MNRAS.373.1074S} and produced highly realistic dwarf galaxies (two classical dwarfs and two faint dwarfs), including their dark matter cores \citep{2014ApJ...789L..17M}. The goal of this work is to investigate whether the superbubble feedback model, which provides a more accurate description of the underlying physics, can produce realistic isolated dwarf galaxies in the $\Lambda$CDM framework, without involving an ad-hoc delayed cooling prescription. In addition, we compare the superbubble model with the blastwave feedback model, and explore in detail how sensitive the formation and properties of dwarf galaxies depends on feedback models.

The paper is organised as follows. In Section~\ref{Sec:sim}, we present the details our simulations, briefly summarizing the main features of the superbubble model. In Section~\ref{Sec:dwarfs}, we describe the two galaxies that formed within our simulation and we compare their properties to the results of previous numerical studies and to the observations. In Section~\ref{subsec:dm}, we study the properties of the host dark matter halos, with particular focus on the cusp--core transformation induced by stellar feedback. In Section ~\ref{sec:dopeyandgrumpy}, we analyse the impact of stellar feedback on the formation of ultra-faint dwarf galaxies. We compare the CGM surrounding the two galaxies to absorption line observations in Section~\ref{sec:cgm}, and present our conclusion in Section~\ref{sec:concl}. 

\begin{figure*}
\includegraphics[width=\linewidth]{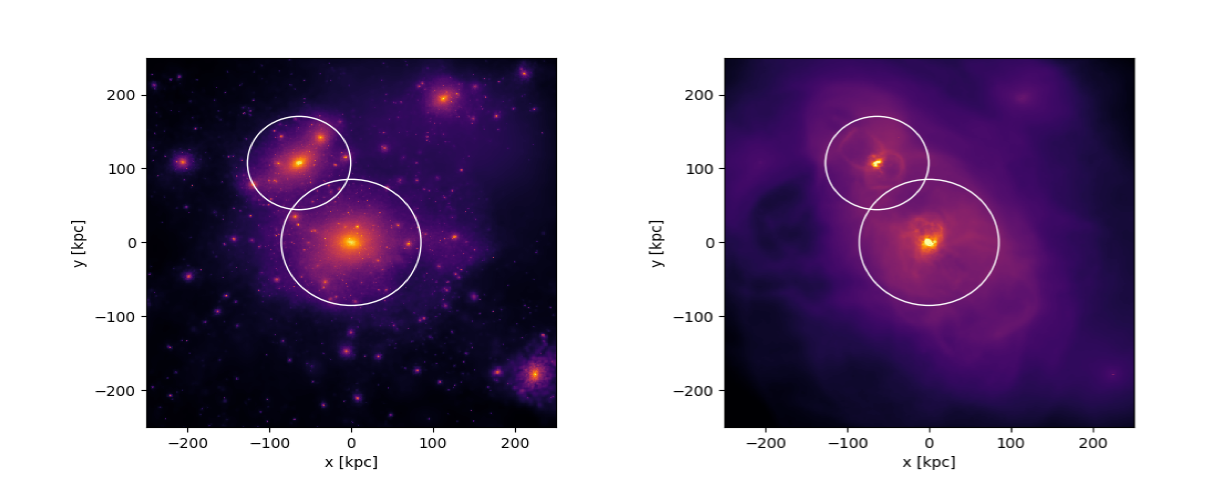} 
\caption{
Projected dark matter (left panel) and gas (right panel) density fields of the simulated dwarfs at $z=0$, in as region of 500 kpc on a side within the zoomed-in region. The projections are centered on Bashful, the most massive among the two simulated dwarfs. The virial radii of the Bashful and Doc are marked by the white circles.
}\label{Fig:dens_z0}
\end{figure*}

\section{Simulation}\label{Sec:sim}
The simulation is performed by using the TreeSPH code \codename \citep{Wadsley2017}, with the same initial conditions and numerical setup as in  \citet{2014ApJ...792...99S}. In brief, the cosmology we adopt is identified by $\Omega_{\rm b} = 0.042$, $\Omega_{\rm m}=0.24$, $\Omega_{\Lambda}=0.76$, $h = 0.73$ and $\sigma_8 = 0.77$. At redshift $z=0$, the simulation box is about $25$ Mpc on a side, with an embedded high resolution region of about $2$ Mpc on a side. The latter contains $6$ millions of dark matter particles and an equal number of SPH particles, with mass resolution of $m_{\rm{dm}} = 1.6 \times 10^4~M_{\odot}$ and $m_{\rm{sph}} = 3.3 \times 10^3~M_{\odot}$, respectively. The gravitational softening length, $\epsilon_{\rm G}$, is set to $86$ pc (physical) from $z=9$ to present, while it evolves as $1/(1+z)$ from the initial redshift of $z=129$ to $z=9$. The smoothing length of SPH particles is allowed to shrink up to 0.1 $\epsilon_{\rm G}$. The simulation includes radiation from a UV background from \citet{2012ApJ...746..125H} and cooling from primordial and metal species over a temperature range of $T=10-10^{9}$ K as described in \citet{Shen2010}.  

The star formation recipe follows the Schmidt law \citep{1959ApJ...129..243S}:
\begin{equation}
\frac{d \rho_{*}}{dt} = 0.1 \frac{\rho_{\rm{gas}}}{t_{\rm{dyn}}}\propto \rho^{1.5},
\end{equation}
where $\rho_{*}$ and $\rho_{\rm{gas}}$ are respectively the stellar and gas densities, while $t_{\rm{dyn}}$ represents the local dynamical time. Star formation occurs stochastically when gas simultaneously reaches temperatures $T<10^4$ K and exceeds a given density threshold of $n_{\rm H} =$ 100 atoms cm$^{-3}$. A single star particle represents a simple stellar population with its own age, metallicity and initial mass function (IMF). Initially, the mass of each star particle is $m_{*} = 10^3 M_{\odot}$ and its IMF is determined according to \citet{2001MNRAS.322..231K}. 
As each stellar population evolves, energy, mass and metals are injected into the ISM through Type Ia and Type II supernovae and stellar winds. While the stellar wind and Type Ia supernova feedback models remain the same as in \citet{2014ApJ...792...99S} as described in \citet{2006MNRAS.373.1074S} (without cooling shutoff), Type II supernova feedback follows the superbubble model described in \citet{2014MNRAS.442.3013K}. In brief, the novelty of this model relies in three main components: thermal conduction, stochastic evaporation of cold gas and sub-resolution multi-phase particles.  

Thermal conduction represents an important process in order to model accurately the evolution of superbubbles. In presence of temperature gradients, the heat flux $Q$ is computed as $Q = - \kappa \nabla T$, where $T$ is the temperature, and $\kappa (T)$ is the conduction coefficient. $\kappa (T)$ can be approximated by $\kappa_{0} T^{5/2} $ in which $\kappa_0 = 6.1 \times 10^{-7}$ $\rm{erg}$ $\rm s^{-1}$ $\rm K^{-7/2}$ $\rm{cm}^{-1}$ in case magnetic fields are not present \citep{1977ApJ...211..135C}. As heat transfer is mediated by electrons, a mass flux from cold to hot phase (i.e. evaporation of cold gas) must also be present in order to not generate currents in the fluid. However, this process takes place on spatial scales which are too small to be resolved. Therefore, \citet{2014MNRAS.442.3013K} introduced a sub-grid treatment for evaporation of cold gas. Given the internal temperature $T$ and the surface area $A$ of the hot bubble, the mass variation is computed according to the following formula:
\begin{equation}\label{eq:mass_flux_1}
	\frac{d M_{\rm b}}{d t} = \frac{4\pi\mu}{25 k_B} \kappa_0 \frac{\Delta T ^ {5/2}}{\Delta x} A ~ ,
\end{equation}
where $\Delta x$ corresponds to the thickness of the boundary layer between cold and hot gas. In practise, we adopt an area estimate of $A=(6\pi h^{2})/N_{\rm hot}$ and $\Delta x = h$, where $h$ is smoothing length of a given hot particle in the bubble and $N_{\rm hot}$ is the number of neighboring hot particles. Cold gas particles are evaporated into the hot phase stochastically by converting Eq.~\eqref{eq:mass_flux_1} into a probability per time-step.  Evaporated gas particles then receive half of the thermal energy of the hot gas particle that causes their evaporation. 

At early stages of the formation of a bubble or, in low resolution simulations, resolution limits often prevent a proper estimation of temperature and density, yielding once again to an overcooling problem. In the superbubble model, rather than shutting down cooling for a given amount of time as done in \citet{2006MNRAS.373.1074S}, fluid elements are allowed to temporarily enter a multi-phase state. Numerically speaking, particles are defined as cold if their temperature falls below a given threshold of $T_{\rm{th}}=10^5$ K. During feedback events, energy is dumped into surrounding gas and particles enter a two-phase state, carrying two values of mass and energy for the hot and the cold phases, which are assumed to be in pressure equilibrium. Thus, each particles have total mass and energy:
\begin{eqnarray}
m &=& m_{\rm{hot}} + m_{\rm{cold}} ~ , \label{eq:tot_m} \\
E &=& u_{\rm{hot}}m_{\rm{hot}} + u_{\rm{cold}}m_{\rm{cold}} ~ , \label{eq:tot_E}
\end{eqnarray}

where $u$ corresponds to the internal energy per unit of mass. By means of Eq. \ref{eq:tot_m} and Eq. \ref{eq:tot_E}, the density of each particle's phase is then computed by assuming pressure equilibrium. $PdV$ work onto multi-phase particles are distributed to the two phases according to their current energy fractions, thus preserving pressure equilibrium. At each time-step, the mass flux between the two phases is determined by the evaporation rate given by Eq.~\ref{eq:mass_flux_1}. More explicitly, 
\begin{equation} \label{eq:mass_flux_3}
    \frac{{d} M_{\rm b}}{d t} = \frac{16 \pi \mu}{25 k_B} \kappa_0 T_{\rm{hot}}^{5/2} h ~,
\end{equation}
where $h$ is the SPH smoothing length. Once the temperature of the hot phase falls below $T_{\rm{th}}$, or the cold phase entirely evaporates, the given particle returns to a single-phase state.

Metal enrichment is modelled in the same way as in \citet{2014ApJ...792...99S}. Basically, the yields of oxygen and iron are tracked separately according to \citet{1996A&A...315..105R}, and converted to alpha-elements and iron-peak elements, respectively, assuming solar abundances patterns \citep{doi:10.1146/annurev.astro.46.060407.145222}. Turbulent mixing of metals (and thermal energy) are modelled with a variant of the \citet{10.1175/1520-0493(1963)091<0099:GCEWTP>2.3.CO;2} turbulent diffusion model, as described in detail in \citet{Shen2010}.

\section{The simulated dwarfs}\label{Sec:dwarfs}

\begin{table*}[t]
\centering
\begin{tabular}{l c c c c c c c c c c c}
\hline\hline 
Name 	& $M_{\rm{vir}}$	& $R_{\rm{vir}}$ 	& $R_{\rm{c}}$ 		& $V_{\rm{max}}$ 			& $M_{*}$ 		& $M_{\rm{gas}}$ 	& $M_{\rm HI}$ 		& $f_{\rm b}$ 				& $\braket{[Fe/H]}$	& $M_{V}$   	& $B-V$ 	\\ [0.5ex] 
		& ($M_{\odot}$)		& (kpc) 			& (kpc) 			& ($\rm{km} ~ \rm{s}^{-1}$) 	& ($M_{\odot}$) 	& ($M_{\odot}$) 	& ($M_{\odot}$) 		& 					& 		    		& 	 		& 		\\ [0.5ex]
\hline
Bashful 	& $3.57\times10^{10}$ & $85.25$ 		& $3.45$ 			& $49.30$ 				& $1.43\times10^{8}$ & $1.27\times10^{9}$ & $7.03 \times 10^7$ 	& 0.039				& $-1.22$			& $-15.49$ 	& $0.66$ \\
Doc 		& $1.45\times10^{10}$ & $63.23$ 		& $1.86$ 			& $39.40$ 				& $1.48\times10^{7}$ & $3.66\times10^{8}$ & $2.72 \times 10^7$ 	& 0.026				& $-1.55$			& $-13.58$ 	& $0.35$ \\
Dopey 	& $3.27\times10^{9}$ & $38.44$ 		& ---				& $22.37$ 				& --- 				 & $4.39\times10^{7}$ & $4.66 \times 10^3$ 	& 0.013				& ---				& --- 		    	& --- \\
Grumpy 	& $1.71\times10^{9}$ & $30.98$ 		& ---				& $21.45$ 				& --- 				 & $9.28\times10^{6}$ & $4.63 \times 10^3$ 	& 0.005	 			& ---				& ---	 	    	& --- \\
\hline
\end{tabular}
\caption{Properties of the simulated dwarf galaxies at the present day. The different columns indicate, respectively, the name of the galaxy, virial mass, virial radius, dark matter core radius, maximum circular velocity, stellar mass, gas mass, \ion{H}{i} mass, baryon fraction $f_{\rm b} \equiv (M_{*}+M_{\rm gas})/M_{\rm vir}$, mean stellar metallicity, V-band magnitude, and $B-V$ color. Note that we define the virial radius as the radius enclosing a mean density of $93$ times the critical density.}
\label{tab:1}
\end{table*}

\begin{figure}
\includegraphics[width=\linewidth]{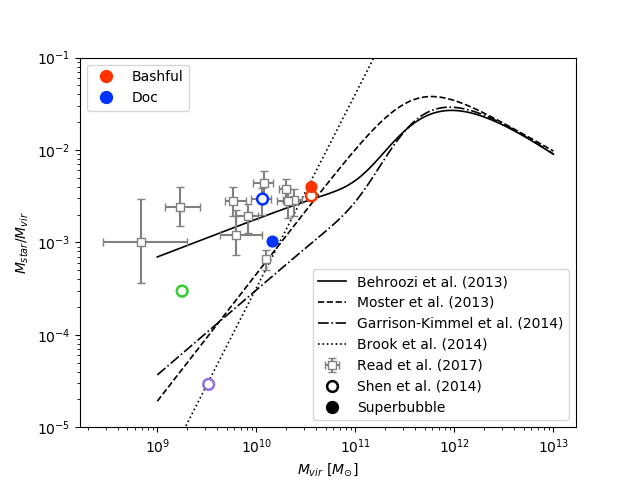}
\caption{
Present day stellar mass fraction of dwarf galaxies. The filled colored circles represent Bashful (red) and Doc (blue) simulated with the supebubble feedback. The empty circles, with colored edges, correspond to Bashful (red), Doc (blue), Dopey (purple) and Grumpy (green) simulated in \shenonex, with the blastwave feedback model. The empty grey squares represent the sample of isolated dwarf galaxies analysed in \citet{2017MNRAS.467.2019R}. The black lines report the low-mass extrapolation of the stellar mass--halo mass relation provided by \citet{2013ApJ...770...57B} (solid), \citet{2013MNRAS.428.3121M} (dashed), \citet{2014ApJ...784L..14B} (dot-dashed), and \citet{2014ApJ...784L..14B} (dotted).
}\label{Fig:smf}
\end{figure}

At the present day, two field dwarf galaxies have formed in our simulation, as we can see in Fig. \ref{Fig:dens_z0}, which shows the integrated dark matter and gas density field. Following the same naming convention as \citet{2014ApJ...792...99S}, the two dwarfs Bashful and Doc inhabit the two most massive dark matter halos found in the high resolution region. 
Other dark halos present in the field, have also accreted gas during the evolution of the simulated Universe. However, in their case, gas was not able to cool enough to ignite star formation.
In particular, smaller dwarf galaxies found in \citet{2014ApJ...792...99S}, such as Dopey and Grumpy, are not able to form here. This suggests that the formation of ultra-faint dwarf galaxies is highly sensitive to external feedback mechanisms, which we discuss in Section~\ref{sec:dopeyandgrumpy}. Some global properties of the four most massive halos in the simulation at z = 0 are listed in Tab.~\ref{tab:1}. All of them are field dwarf galaxies with the nearest massive halo of $M_{\rm vir} = 2.5 \times 10^{12} ~M_{\odot}$ are more than $3$ Mpc away.

\subsection{Stellar mass-halo mass relationship}
 At redshift $z = 0$, Bashful and Doc have stellar masses of $M_{*} = 1.4 \times 10^{8}~M_{\odot}$ and $M_{*} = 1.5 \times 10^{7}~M_{\odot}$, corresponding to stellar mass fractions $M_{*}/M_{\rm{vir}}$ of 0.004 and 0.001, respectively. Fig. \ref{Fig:smf} shows the stellar mass-halo mass relation (SMHM) for the two dwarfs, together with abundance matching results from \citet{2013ApJ...770...57B} and \citet{2013MNRAS.428.3121M}, and more recent relations derived from low mass Local Group dwarf galaxies \citep{2014ApJ...784L..14B,2014MNRAS.438.2578G}, and results from \citet{2017MNRAS.467.2019R} where the stellar mass of isolated dwarf galaxies are derived from fitting their rotation curves. As shown in Fig. \ref{Fig:smf}, the stellar mass fractions of Bashful and Doc are in excellent agreement with abundance matching results from \citet{2013MNRAS.428.3121M} and \citet{2014ApJ...784L..14B}. Both Bashful and Doc appears slightly above the stellar-to-halo mass relations of \citet{2014MNRAS.438.2578G}, but Doc lies below the \citet{2013ApJ...770...57B} relation and the data points form \citet{2017MNRAS.467.2019R}. Given the large uncertainties associated with the SMHM relation at low-mass regimes, our results are in general well consistent with abundance matching results. Comparing to the results from the simulations with the blastwave feedback in \citet{2014ApJ...792...99S}, Bashful has a similar stellar mass, whereas Doc's stellar mass has reduced by more than a factor of $2$, indicating that the superbubble feedback is more effective regulating star formation for low-mass galaxies. 

We used the Padova simple stellar populations models \citep{2008A&A...482..883M,2010ApJ...724.1030G} to compute broadband luminosities. Bashful and Doc have V-band magnitudes of $M_{V} =-15.49$ and $-13.58$, respectively. If they were to be observed today, with values of $B-V$ of $0.66$ for Bashful and $0.35$ for Doc, they would appear blue and they would be classified as dwarf irregulars (dIrrs). 
At $z = 0$, in our simulation volume, the two dwarfs are located within $160$ kpc from each other.

\subsection{Star formation history}

\begin{figure*}[ht]
\centering
\includegraphics[width=\linewidth]{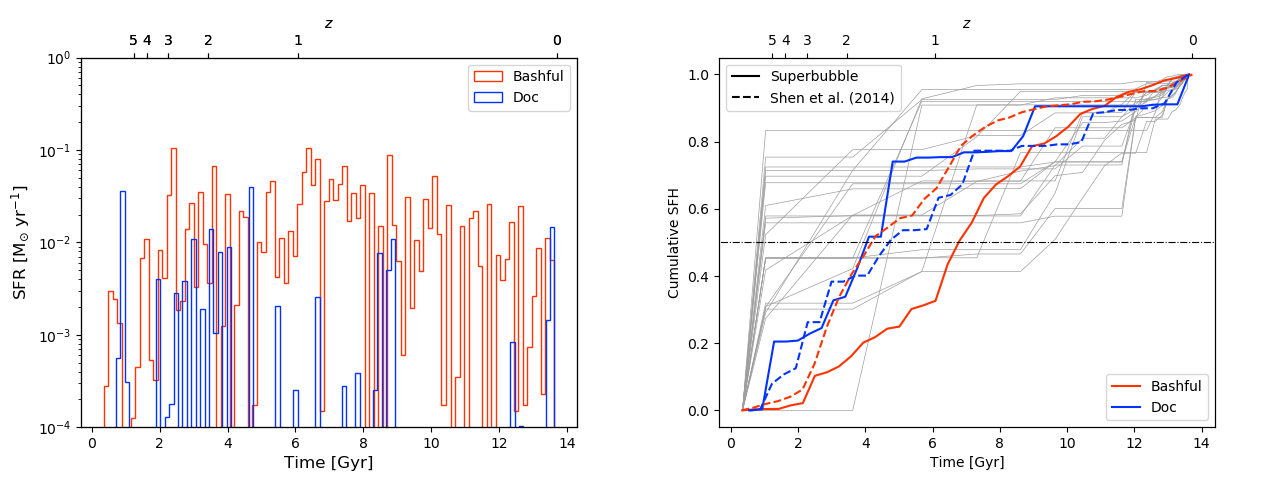}
\caption{
Left panel: star formation history of Bashful (red) and Doc (blue), averaged a over time intervals of $140$ Myr. Right panel: cumulative star formation history of the simulated dwarf galaxies, with total stellar mass normalised to unity. The solid colored lines represent Bashful (red) and Doc (blue) in our simulation, while the dashed colored lines correspond, with the same colors, to the dwarfs galaxies simulated in \shenonex. The gray solid lines are the cumulative star formation histories of individual dwarf irregulars in the ANGST sample \citep{Weisz_2011}.
}\label{Fig:SFH}
\end{figure*}

\begin{figure*}[ht]
\centering
\includegraphics[width=\linewidth]{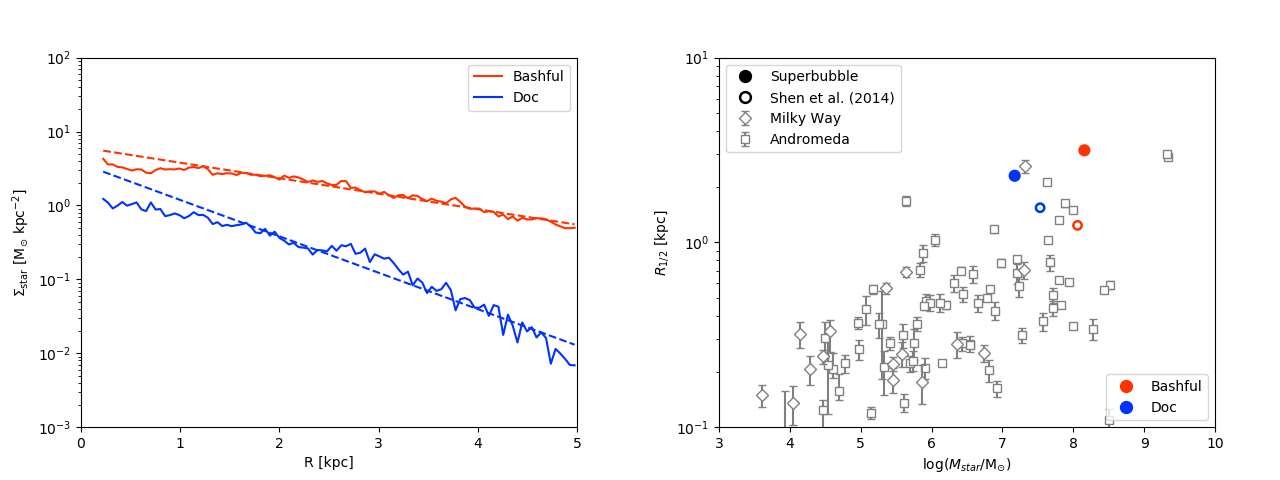}
\caption{
Left panel: 1D line-of-sight stellar surface density profiles of Bashful (red) and Doc (blue), represented by the solid lines. The dashed lines show the exponential fit of their stellar disks, $I(R>1~{\rm kpc}) = I_{\rm d} \exp(-R/R_{\rm d})$, with a characteristic scale of $R_{\rm d} = (2.07 \pm 0.036)$ kpc for Bashful and $R_{\rm d} = (0.88 \pm 0.023)$ kpc for Doc. Right panel: stellar half-light radius-mass relation of the simulated dwarfs at $z=0$. The filled colored circles represent Bashful (red) and Doc (blue) in our simulation. 
The empty circles, with colored edges, correspond to Bashful (red) and Doc (blue) simulated in \shenonex, with the blastwave feedback model. 
The other empty symbols represent satellite galaxies of the Milky Way (diamonds) and Andromeda (squares), from the original sample of Local  Group galaxies published by \citet{2012AJ....144....4M}.}\label{Fig:stars}
\end{figure*}

In the left panel of Fig.~\ref{Fig:SFH}, we plot the star formation histories (SFHs) of the two simulated dwarfs. Both Bashful and Doc present an extended SFH, characterised by stochastic bursts, each followed by a period of quiescence. Although photo-heating from reionisation suppresses gas accretion, for halos with masses of Bashful and Doc, the primordial gas is able to cool down its temperature via Compton and radiative cooling and condense at the center of the gravitational potential well. As gas cools further, stars are ignited and, along with the first bursts, SN-driven outflows deplete the central regions of star-forming material. It follows a short quiescence phase, during which cold gas is accreted towards the center and a new cycle starts again. Bursty star formation is commonly seen in simulations with explicit, strong supernova feedback models \citep[e.g.,][]{2012MNRAS.422.1231G,Teyssier2013,2014ApJ...792...99S,10.1093/mnras/stv2072,2017MNRAS.471.3547F,2019MNRAS.482.1176W}, which often leads to rapid change of inner gravitational potential wells and to a flattening of central dark matter profiles \citep[e.g.,][]{2010Natur.463..203G,2012MNRAS.421.3464P,2014ApJ...789L..17M}. DM core formation is also seen in our simulation, and we discuss this in more detail in Section ~\ref{subsec:dm}. 

While the SFH of Bashful shows no sign of long interruption periods, Doc shows signs of strong suppression of star formation over longer period of time. It completely stops forming stars for a $\gtrsim 3$ Gyr period between $0.5<z<0.1$ and, after that, it starts forming stars again. Investigations of resolved stellar populations in the nearest dwarf galaxies found that bursts of star formation can be separated in time by tens or hundreds of Myr to a few Gyr \citep[e.g.,][]{2009ARA&A..47..371T}.
This result is similar to that in \citet{2019MNRAS.482.1176W}, where the authors simulated dwarf galaxies in the mass range of $M_{\rm vir} \sim 10^{9-10}~\text{M}_{\odot}$ and found that a large fraction of them ($\sim 20$\%) show at least one long period of quiescence. 
In their work, the dwarfs resuming start formation after a long period of quiescence tends to have a higher-than-average ratio of $M_{\rm HI}/M_{\rm *}$, and they started forming stars again after experiencing interaction with streams or filaments of gas from the cosmic web or nearby mergers. This appears to be also the case in Doc, becasue the whole system is merging towards $z=0$.

In the right panel of Fig.~\ref{Fig:SFH}, we plot the cumulative star formation histories of the two dwarfs. For comparison, we plot in the same figure cumulative star formation histories of individual dIrrs found in the ACS Nearby Galaxy Survey Treasury (ANGST) sample, previously analysed in \citet{Weisz_2011}, together with those of the dwarfs found in \shenonex. Both Bashful and Doc are actively forming stars today, which is consistent with observations of field dwarf galaxies. For instance, a fraction of $80$\% of galaxies in the ANGST catalogue are actively forming stars, and in  \citet{2012ApJ...757...85G}, they found that field dwarfs with a stellar mass in the range $10^7 M_{\odot} < M_{*} < 10^9 M_{\odot}$ which are not forming stars today are extremely rare. In \citet{Weisz_2011}, it was found that, on average, $50$\% of the stellar mass of a dwarf galaxy is formed by redshift $z \sim 2$, while $60$\% prior to redshift $z \sim 1$. As in the case of \shenonex, both Bashful and Doc have SFHs highly compatible with those in the ANGST catalogue. 

 In our work, Doc exhibits a similar trend to the corresponding dwarf in \shenone in terms of cumulative star formation history, even if it exhibits fewer and generally smaller star formation peaks, and longer periods of quiescence compared to the case of the blastwave model. However, early star formation in Bashful appears to be more suppressed with the superbubble feedback, such that it forms half of its stellar mass by $z\lesssim 1$, about 3 Gyr later than Bashful in \citet{2014ApJ...792...99S}. This result, together with the fact that Doc has less stellar mass and longer quiescent periods, indicate that the superbubble feedback is more effective for low mass galaxies with $M_{*} \leq 10^{7}$ M$_{\odot}$, although we note that the SFR in low mass galaxies like these may be subject to strong stochastic effects \citep{Keller_2019}.

\hfill
\newline
\subsection{Stellar distribution and kinematics}

\begin{figure}
\includegraphics[width=\linewidth]{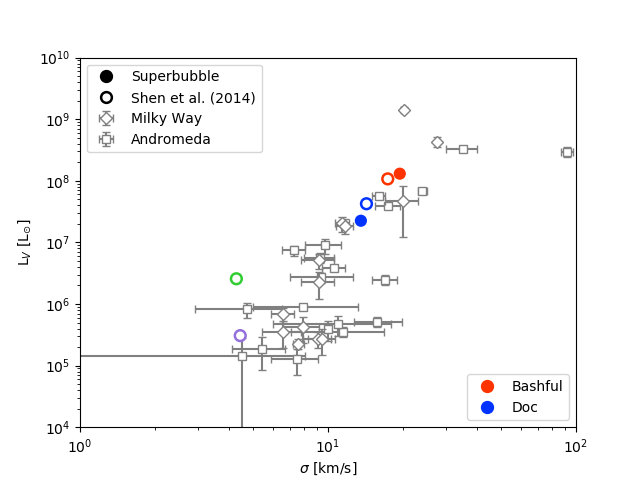}
\caption{
V-band luminosity versus stellar 1D line-of-sight velocity dispersion of the two simulated dwarfs. The V-band luminosity is computed by using the Padova simple stellar populations models \citep{2008A&A...482..883M,2010ApJ...724.1030G}. The filled colored circles represent Bashful (red) and Doc (blue) in our simulation. The empty circles, with colored edges, correspond to Bashful (red), Doc (blue), Dopey (purple) and Grumpy (green) simulated in \shenonex, with the blastwave feedback model. The empty symbols show the observed $L_V-\sigma$ relation for satellite galaxies of the Milky Way (diamonds) and Andromeda (squares), from the original sample of Local Group galaxies published by \citet{2012AJ....144....4M}.
}\label{Fig:r_2sm}
\end{figure}

Both Bashful and Doc exhibit exponential stellar disks at the present day. Solid lines in Fig.~\ref{Fig:stars} (left panel) correspond to 1D line-of-sight stellar surface densities as a function of the radial distance from the center for Bashful and Doc. The dashed lines correspond to exponential fits ($I(R>1~\text{kpc}) = I_{\rm d} \exp \left ( -R/R_{\rm d} \right )$) to their stellar disks, at radii between $1< R <5$ kpc. 
The scale length of the two stellar disks are $R_{\rm d} = (2.07 \pm 0.036)$ kpc and $R_{\rm d} = (0.88 \pm 0.023)$ kpc, for Bashful and Doc, respectively. As in \shenonex, the two dwarfs show no sign of central bulge and the stellar disks extend to $5$ kpc from the center. There is a signature of a stellar core also in Doc's stellar density profile, similar to that in \citet{2014ApJ...792...99S}, which is also seen in the observed dIrr galaxy WLM \citep{2012ApJ...750...33L}. It is worth noting that the scale length of Bashful is more than twice as the one in \citet{2014ApJ...792...99S} with the blastwave feedback ($R_{\rm d}= 0.97$ kpc). One possible explanation is that star formation in Bashful is shifted to later time with the superbubble feedback. The ISM and stellar populations are generally more compact in high redshift galaxies due to more frequent merger events, and the suppression of high-z star formation activities leads to larger disks \citep{2012MNRAS.426..690B}. Moreover, similar to dark matter particles, stars are also collisionless particles. Thus, the same mechanism which causes central dark matter particles to migrate outwards (and thus creates dark matter cores) is also responsible for dynamical heating of the stellar disk. As we show in Section \ref{subsec:dm}, superbubble feedback also leads to rapid fluctuations of the central potential well, resulting in a twice-as-large DM core in Bashful. The strong correlation between DM core size and stellar disk scale length is consistent with previous suggestions by \citet{2012MNRAS.421.3464P}, and also with observations \citep{2009Natur.461..627G}.

In the right panel of Fig.~\ref{Fig:stars}, we plot instead the stellar half-light radius, denoted by $R_{1/2}$, versus the stellar mass of the simulated dwarfs, at $z=0$. Bashful and Doc have $R_{1/2}$ of $3.15$ and $2.29$ kpc, respectively, both are consistent (although on the higher side) with 
satellites of the Milky Way and Andromeda galaxies \citep{2012AJ....144....4M}. Similar to the disk scale lengths, $R_{1/2}$ from superbubble feedback appears to be significantly larger than the ones in \citet{2014ApJ...792...99S} with the delayed cooling model,  especially in the case of Bashful. This is again due to the late star formation history and stronger dynamical heating from the superbubble feedback, as discussed before.

In Fig.~\ref{Fig:r_2sm}, we show the V-band luminosity of the simulated dwarfs against their 1D line-of-sight (LOS) stellar velocity dispersion ($\sigma_{*}$), together with the observational data of the dwarf galaxies within the Local Group compiled in \citet{2012AJ....144....4M}. The velocity dispersion for each galaxy was computed along 3 orthogonal lines of sight for all stars within 2 disk scale lengths ($R_{\rm d}$). The $\sigma_{*}$ of Bashful and Doc are 20.8 km/s and 14.8 km/s, respectively, and the $L_V$--$\sigma_{*}$ relations of Bashful and Doc appear highly consistent with the observations. There is no systematic difference in stellar kinematics between the superbubble and blastwave models.
As in \citet{2014ApJ...792...99S}, we also found that $\sigma_{*}$ increases with stellar age. Stars are born kinematically cold and heated by galaxy mergers and outflows. As larger galaxies often form earlier, have higher star formation rates, and undergo more merger events, it is reasonable to expect that they have larger velocity dispersions. It has also been shown that for satellite galaxies, tidal stripping can decrease $\sigma_{*}$ significantly when dIrrs accrete onto larger galaxies and are transformed into dSphs \citep[e.g.,][]{Lokas_2011,Brooks_Zolotov_2014}, but this process is less relevant for field dwarf galaxies like in this work. We also measured the stellar rotational velocities ($v_{rot,*}$) by taking the azimuthal velocities at $2 R_{\rm d}$, and found that $v_{\rm{rot},*}$ is 11.1 km/s and 13.6 km/s for Bashful and Doc, respectively. Thus, the ratio between rotational velocity and velocity dispersion, $(v_{\rm rot}/\sigma)_{*}$, is 0.53 and 0.92 for Bashful and Doc, respectively, which indicates that the stellar disks are pressure-dominated ($(v_{\rm rot}/\sigma)_{*} \la 1$). This is also consistent with observations of faint dIrrs ($v_{\rm rot} \la 20$ km/s) in the Local Volume, where a large fraction of them are found dispersion-dominated.

\subsection{Cold Gas Properties}
\begin{figure}
\includegraphics[width=\linewidth]{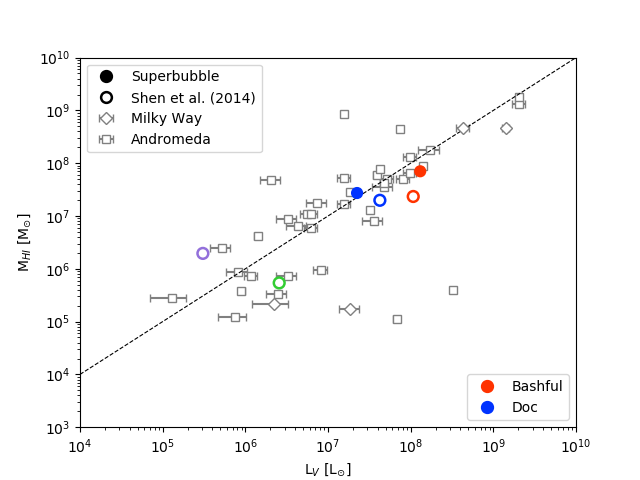}
\caption{
Total neutral hydrogen (\ion{H}{i}) mass versus V-band luminosity for the simulated dwarf galaxies. We compute the \ion{H}{i} mass directly from the simulation, where the abundance of \ion{H}{i} is calculated by integrating the ionisation equations for primordial species. The filled colored circles represent Bashful (red) and Doc (blue) in our simulation, at $z=0$. 
The empty circles, with colored edges, correspond to Bashful (red), Doc (blue), Dopey (purple) and Grumpy (green) simulated in \shenonex, with the blastwave feedback model. The empty symbols represent satellite galaxies of the Milky Way (diamonds) and Andromeda (squares), from the original sample of Local Group galaxies published by \citet{2012AJ....144....4M}. The dashed line indicates $L_{V}/M_{\rm HI}$ =1. 
}\label{Fig:gas_lum}
\end{figure}

As listed in Tab. \ref{tab:1}, at $z=0$, Bashful and Doc has total mass in neutral hydrogen $M_{\rm HI} = 7.03 \times 10^{7} M_{\odot}$ and $2.72 \times 10^{7} M_{\odot}$, respectively. Both values are approximately twice as much gas mass compared to the corresponding dwarfs in \citet{2014ApJ...792...99S} with the blastwave supernova feedback model, although their stellar mass are comparable or smaller. It appears that the superbubble feedback is able to retain more cold gas within the galaxies while regulating star formation, and this can be seen in Fig.~\ref{Fig:gas_lum}, where we plot the relation between \ion{H}{i} gas mass and V-band luminosity for the two simulated dwarfs, together with the data sample published in \citet{2012AJ....144....4M}. Bashful and Doc appear to be consistent with the observed dwarfs in the Local Group. The total amount of cold gas increase with stellar luminosity, because more massive galaxies can sustain galactic fountains and re-accrete gas previously blown away by supernova feedback, and cold gas is also less susceptible to photoionisation heating from the UV background. Nevertheless, it is found that the ratio of \ion{H}{i} gas mass over stellar mass are generally decreasing with increasing stellar mass, i.e. smaller galaxies are more gas rich. For example, the ALFALFA extragalactic H I survey find more than half of their sample has $M_{\rm HI}/M_{*} > 1$, in particular galaxies with $M_{*} \lesssim 10^{7}~M_{\odot}$ \citet{2012AJ....143..133H}. The \ion{H}{i} gas mass to stellar mass ratio $M_{\rm HI}/M_{*}$ are 0.49 and 1.83 for Bashful and Doc, respectively, which are in excellent agreement with the ALFALFA observations. 

In contrast to stellar kinematics, the cold gas disks of these systems appear to be rotationally supported. At z = 0, Bashful and Doc have HI velocity dispersion ($\sigma_{\rm HI}$) of 11.8 km/s and 4.9 km/s, respectively, in good agreement with observed $\sigma_{\rm HI}$ for dIrrs in the Local Group \citep{Stilp2013}. The $v_{\rm rot}/\sigma$ ratio for the HI disk is 1.78 for Bashful and 3.52 for Doc, significantly higher than the ratios for the stellar disks. Interestingly, the $v_{\rm rot}/\sigma$ ratio for cold gas is larger in our simulation than in \citet{2014ApJ...792...99S} with the delayed cooling model. The HI gas has higher rotational velocity and lower velocity dispersion, suggesting that the cycle of baryons modulated by the superbubble feedback retains more high angular momentum gas, and generates less perturbations in the ISM. We defer a more detailed study of the angular momentum of the accreting gas, outflows, the CGM and the ISM in a future work.

We note that Dopey and Grumpy also have a small amount of neutral hydrogen, of order a few times $10^{3}~M_{\odot}$, which is significantly less than the total gas mass within the halo (about $10^{7}~M_{\odot}$), indicating most gas is in ionised form. Unlike in \citet{2014ApJ...792...99S}, gas in these two halos is generally unable to cool and fuel star formation due to the influence from the neigbouring more massive halos. We discuss this further in Section \ref{sec:dopeyandgrumpy}.

\subsection{Mass-metallicity relationship}\label{sec:mzr}

\begin{figure*}
\includegraphics[width=\linewidth]{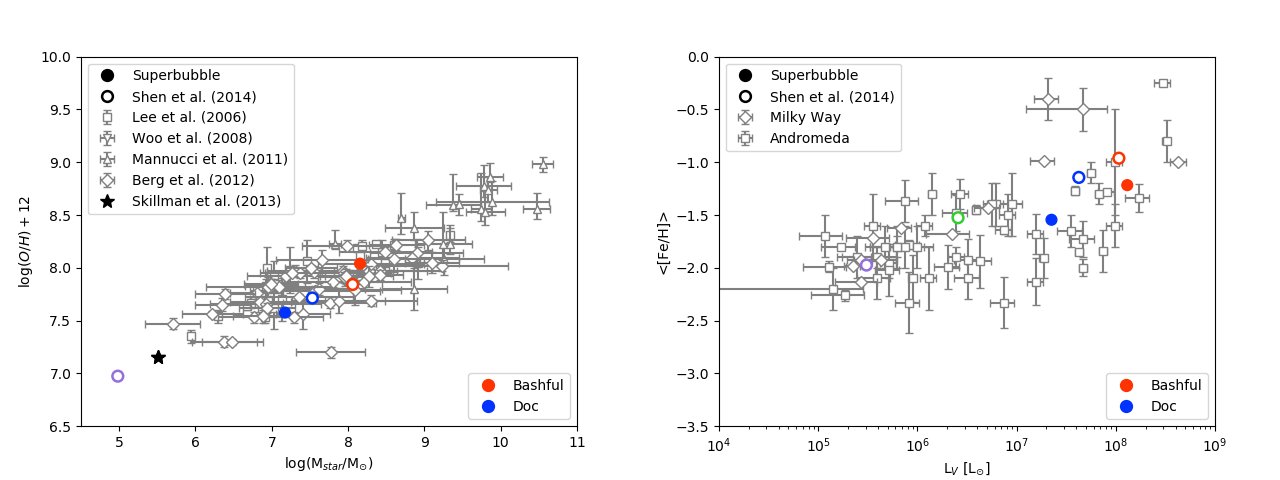}
\caption{
Left panel: stellar mass-gas phase metallicity relation of low-mass galaxies. The filled colored circles represent Bashful (red) and Doc (blue) in our simulation, at $z=0$. The empty circles, with colored edges, correspond to Bashful (red), Doc (blue), Dopey (purple) and Grumpy (green) simulated in \shenonex, with the blastwave feedback model. Other symbols show the observed stellar mass-gas phase metallicity relation for nearby dwarf irregulars \citep[empty squares][]{Lee_2006}, Local Group dwarfs \citep[empty triangles down][]{10.1111/j.1365-2966.2008.13770.x}, host galaxies of $z < 1$ long gamma-ray bursts \citep[empty triangles up,][]{10.1111/j.1365-2966.2011.18459.x}, low--luminosity galaxies in the local volume \citep[empty diamonds][]{Berg_2012}, and Leo P \citep[black filled star][]{Skillman_2013}.
Right panel: stellar metallicity versus V-band luminosity for the simulated dwarf galaxies at $z=0$. Our simulated dwarf galaxies and those simulated in \shenone are represented with the same colors and symbols as in right panel. The empty symbols represent satellite galaxies of the Milky Way (diamonds) and Andromeda (squares), from the original sample of Local Group galaxies published by \citet{2012AJ....144....4M}.
}\label{Fig:metals}
\end{figure*}

In the left panel of Fig.~\ref{Fig:metals}, we plot the gas-phase oxygen abundance versus stellar mass for the two simulated dwarfs. In the same figure, we also show the mass-metallicity relation observed over several orders of magnitude in stellar mass, from the brightest spirals to dIrrs \citep{Lee_2006,10.1111/j.1365-2966.2008.13770.x,10.1111/j.1365-2966.2011.18459.x,Berg_2012,Skillman_2013}. Note that, the observed mass-metallicity relation holds only for star forming galaxies, as gas-phase oxygen abundances are inferred from emission lines due to photo-ionization processes from young stars. Also the global shape of the curve strongly depends on the empirical calibration used and, as a consequence, the gas-phase oxygen abundances can differ up to 0.7 dex, as shown by \citet{2008ApJ...681.1183K}. The gas-phase oxygen abundances estimated for Bashful and Doc follow the measured mass-metallicity relation remarkably well.

In the right panel of Fig.~\ref{Fig:metals}, the average metallicity of Bashful and Doc are plotted and compared together with the V-band luminosity-metallicity relation measured for the dSph satellites of the Milky Way, by 
\citet{2012AJ....144....4M}, which spans several dex in luminosity. Although the very low luminosity tail of the observed V-band--metallicity relation, corresponding to ultra-faint dwarf galaxies, exhibits a lot of scatter, the luminosity range which brackets our two dwarfs is well constrained. 
The low luminosity of low mass galaxies is attributed to the old stellar populations, as opposed to higher mass systems where the majority of stars formed more recently.

\section{Dark Matter}\label{subsec:dm}

\begin{figure}[ht]
\centering
\includegraphics[width=\linewidth]{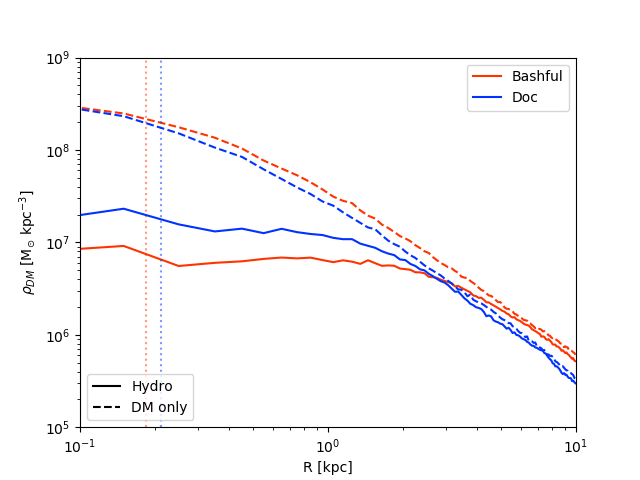}
\caption{
Present day dark matter density profiles of Bashful (red) and Doc (blue). The solid lines represent the hydrodynamic simulation, while the dashed lines correspond to the DM only control run. The dotted vertical lines denote the radius within which numerical convergence is not achieved, due to two-body relaxation \citep{2003MNRAS.338...14P}.
}\label{Fig:dm_dens_prof}
\end{figure}

\begin{figure*}[ht]
\centering
\includegraphics[width=\linewidth]{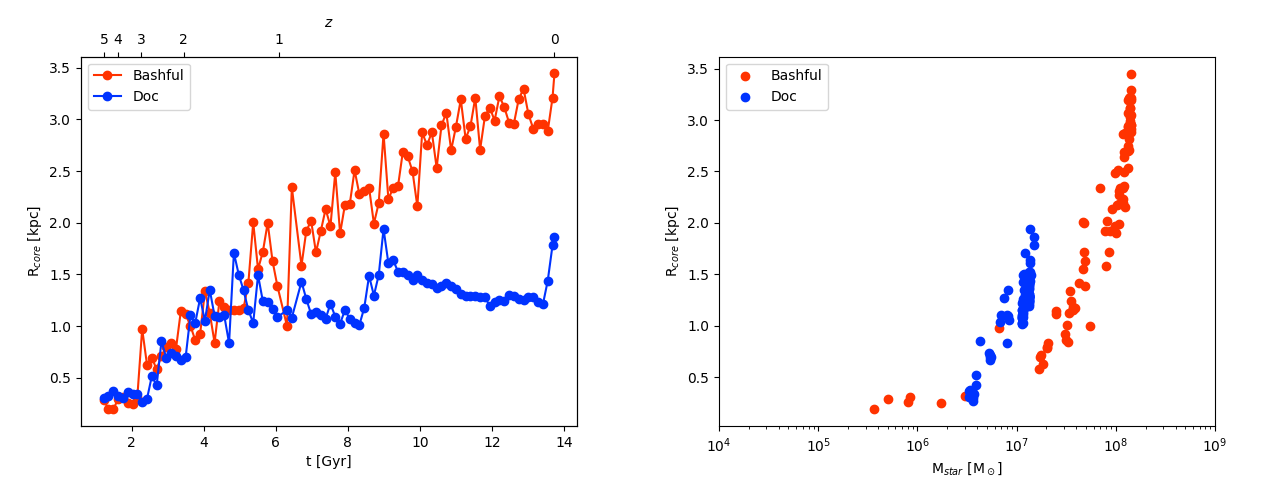}
\caption{
Left panel: temporal evolution of the core size for the two dwarfs. The rich stellar activity of Bashful (red) leads to a central core which consistently grows in time until it reaches a present day core radius of $R_{\rm c} = 3.45$ kpc. During long quiescence periods, the dark matter core of Doc (blue) remains stable, albeit with a slow decrease in size, and its the present day size is $R_{\rm c} = 1.86$ kpc. Right panel: evolution of the core radius for Bashful (red) and Doc (blue), as a function of their stellar masses.
}\label{Fig:dm_core}
\end{figure*}

\begin{figure}
\includegraphics[width=\linewidth]{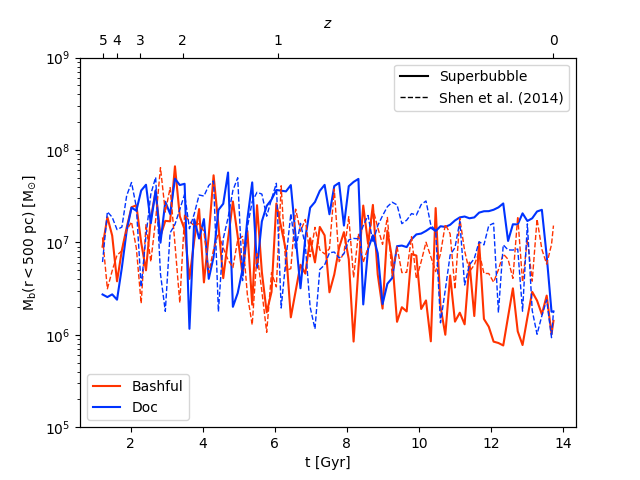}
\caption{
Temporal evolution of the total baryonic mass enclosed within $500$ pc from the center of the two simulated dwarfs. The solid lines correspond to Bashful (red) and Doc (blue) in our simulation, with the superbubble feedback model. With the same color convention, the same two dwarfs analysed in \shenone and \shentwo are represented with dashed lines.
}
\label{fig:Mb500_time}
\end{figure}

The inner structure of dark matter halo hosting dwarf galaxies, at the present day, is different from what was expected from early predictions. In many studies, such as \citet{Walker_2011}, \citet{Adams_2014}, \citet{10.1111/j.1365-2966.2011.19684.x}, \citet{Agnello_2012}, and \citet{Oh_2015}, it has been claimed that observational data favours shallow dark matter density profiles in the innermost region, in form of cores. Despite halos in dark matter only simulations exhibit density profiles well fitted by a universal NFW profile \citep{1996ApJ...462..563N}, recent works have shown how repeated and violent bursts of supernova explosions can significantly affect the central gravitational potential wells of dwarf galaxies in short time scales, which transfers kinetic energy from gas to dark matter particles and lead to cusp-core transformation \citep[e.g.,][]{10.1111/j.1365-2966.2004.08424.x, 2010Natur.463..203G, 2012MNRAS.422.1231G,2012ApJ...759L..42P, 2012MNRAS.421.3464P,2014ApJ...789L..17M,10.1093/mnras/stv2072,2017MNRAS.471.3547F}.

In Fig. \ref{Fig:dm_dens_prof}, we plot the present day dark matter density profiles for Bashful and Doc and we compare them with the corresponding profiles in the DM-only control run. Core formation is observed in the dark matter profiles for both galaxies even at high redshift. The profiles are well fitted by the pseudo--isothermal fitting formula \citep{Oh_2008}: 
\begin{equation}
\rho_{\rm{DM}}(r) = \frac{\rho_0}{1 + (R/R_c)^2} ~ ,
\end{equation}
where $\rho_0$ and $R_c$ are free parameters and they correspond to the central density and the core radius, respectively. 

The present day core sizes of Bashful and Doc are $R_{\rm c} = (3.45 \pm 0.082)$ kpc and $R_{\rm c} = (1.86 \pm 0.032)$ kpc, respectively. The core size of Bashful is twice as large as the one reported in \citet{2014ApJ...789L..17M} ($1.77$ kpc), although it only has about 20\% more stellar mass. Likewise, Doc has a DM core with similar size as in \citet{2014ApJ...789L..17M} ($2.07$ kpc) but its total stellar mass is only about half of that in \citet{2014ApJ...789L..17M}. The differences in dark matter mass within $3.5$ kpc between the dark matter-only run and the hydrodynamic run, $\Delta M_{\rm DM}$, are $5.79\times 10^8~M_{\odot}$ and $2.89\times 10^8~M_{\odot}$ for Bashful and Doc, respectively. And the ``DM core removal efficiency'', defined as $\Delta M_{\rm DM}/M_{*}$, are $6.55$ for Bashful and $19.39$ for Doc, significantly higher than that in \citet{2014ApJ...789L..17M}. This suggests that the superbubble feedback is more effective in creating DM cores. 

The evolution of $R_c$ as a function of time (redshift) is shown in the left panel of Fig.~\ref{Fig:dm_core}. Similar to \citet{2014ApJ...789L..17M}, core formation starts early, because unbinding of the cusp is less energetically demanding when the halo has a shallower potential well. More quantitatively, the minimum energy required for the cusp-core transformation was estimated in \citet{2012ApJ...759L..42P} as $\Delta W/2 \equiv (W_{\rm{core}}-W_{\rm{cusp}})/2$, where $W$ is the gravitational binding energy of the dark matter halo, 
\begin{equation}
    W = -4 \pi G \int_{0}^{R_{\rm{vir}}} \rho_{DM} M_{\rm{enc}} R ~ dR ~ .
\end{equation}
Here, $\rho_{\rm{DM}}$ is the dark matter density, while $M_{\rm{enc}}$ corresponds to the total mass enclosed in a sphere of a radius $R$. We estimated $\Delta W/2$ and compare this with the total energy injected by Type II supernova ($\Delta E_{\rm SN}$), and find that $\Delta E_{\rm SN} > \Delta W/2$ at all times for Bashful and Doc, this again is in agreement with results from \citet{2014ApJ...789L..17M}. For example, at $z=0$, $(\Delta E_{\rm SN}, \Delta W/2) = (2.6 \times 10^{57} \ \rm erg, 4.7 \times 10^{56} \ \rm erg)$ for Bashful and $(2.7 \times 10^{56} \ \rm erg, 1.1 \times 10^{56} \ \rm erg)$ for Doc.
By $z=2$, DM cores of size $\sim 1$ kpc have already been in place for both galaxies. At a given redshift, the size of the core and the central density of dark matter halos is strongly correlated with their prior star formation history. In this case, the rich stellar activity of Bashful, with frequent and continuous bursts, allows its core to grow in size until the present day. 
During long periods of quiescence, the DM core of Doc remains stable, indicating the core-formation process is largely irreversible, as suggested in \citet{2012MNRAS.421.3464P}. However, the core size does decrease slightly, possibly due to new infalling material with high phase-space density. However, this process takes place over a time scale much longer than the core growth timescale induced by stellar bursts, as shown in left panel of Fig.~\ref{Fig:dm_core}.

In the right panel of Fig.~\ref{Fig:dm_core}, we plot the evolution of the core radius as a function of the stellar masses of the two simulated dwarfs. Stellar masses of $M_{*} \gtrsim 10^6 M_{\odot}$ are enough to produce a central cores as large as $\sim 1$ kpc, which is similar to what is found with the blastwave model \citet{2014ApJ...789L..17M}, and is in good agreement with the results of \citet{10.1093/mnras/stv2072} and \citet{2017MNRAS.471.3547F}.
In general, dark matter halos with lower masses have also lower stellar contents, resulting in an insufficient energy input from stellar feedback to relax dark matter density cusps into shallower central densities \citep{2012MNRAS.422.1231G, 2014MNRAS.437..415D, 2014ApJ...789L..17M, 10.1093/mnras/stv2072}. The minimum stellar mass required for the cusp--core transformation and the typical core sizes appear to vary depending on the feedback models employed. The amplitude fluctuations of the central gravitational potential depend upon the dynamics of the large-scale galactic outflows, which is determined by how the underlying feedback model operates. 
The existence of a transition stellar mass, below which stellar feedback results ineffective in forming dark matter cores, is related to the strong dependence of the $M_{\rm *}$ and the $M_{\rm vir}$, as halos with only tiny differences in virial masses can vary by orders of magnitude in stellar content.
For example, in \citet{2016MNRAS.457.1931S} and \citet{2018A&A...616A..96R} no dark matter core was found beyond the gravitational softening length of particles, while in \citet{2016MNRAS.459.2573R} dark matter cores were found at all probed halo masses.

To investigate why the superbubble feedback is more effective creating DM cores, we plot the temporal evolution of total baryonic mass within $500$ pc from the center for both runs in Fig.~\ref{fig:Mb500_time}. Energetic bursts of star formation shown in Fig.~\ref{Fig:SFH} are closely tracked by the cyclic behaviour of the total baryonic mass within $500$ pc from the center of Bashful and Doc. The fluctuations in the central baryonic mass found for our simulated dwarfs have very similar amplitude to the ones found in \shentwox, but in general the superbubble feedback is more effective in quenching star formation.
By explicitly modelling the thermal conduction between the hot and the cold phase of feedback particles, the superbubble model describes better the conversion of thermal energy injected by SNe into kinetic energy. This results into a higher coupling between the energy released by SNe and the ISM. 
This explains why, especially for Doc, in our simulation the total amount of gas mass converted into stars is generally lower than in \shentwo at a given redshift, but the fluctuations in the total baryonic content within $500$ pc from the center of the simulated dwarfs is very similar in the two cases. 
In addition, in our simulation Bashful have a present day stellar mass similar to the corresponding dwarf in \shentwox. However, Fig.~\ref{Fig:SFH} shows that the stellar activity of Bashful is distributed within the entire life of the galaxy and its dark matter core consistently grows until the present day.
Instead, in \shentwo the stellar activity of the same dwarf is more concentrated within the first $\sim 9$ Gyr and, as a consequence, the dark matter core of Bashful does not grow much after $9$ Gyr. However, it is worth nothing that stochasticity in numerical simulations may alter the SFH \citep[e.g., ][]{Keller_2019} and hence core formation, growth and final size.

\section{What killed Dopey and Grumpy?} \label{sec:dopeyandgrumpy}

\begin{figure}
\centering
\includegraphics[width=\linewidth]{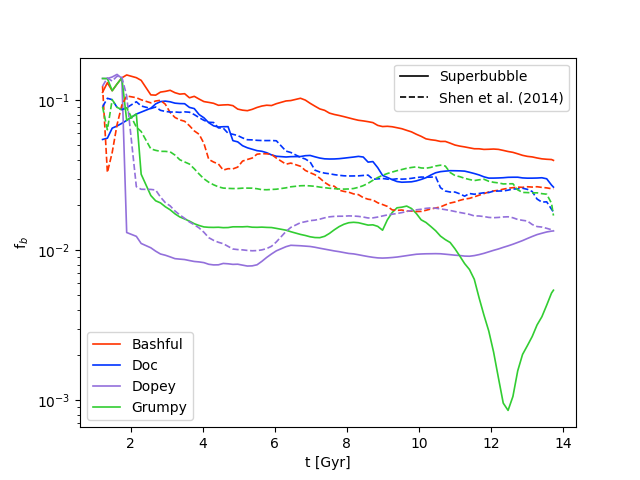}
\caption{
Temporal evolution of the baryonic fraction $f_{\rm b} \equiv (M_{*} + M_{\rm gas})/M_{\rm vir}$ of the Bashful (red), Doc (blue), Dopey (purple) and Grumpy (green). The solid lines correspond to the dwarfs simulated with the superbubble feedback, while the dashed lines correspond to the dwarfs simulated in \shenone with the blastwave feedback.
}\label{Fig:baryonfrac_evo}
\end{figure}

\begin{figure*}
\includegraphics[width=\linewidth]{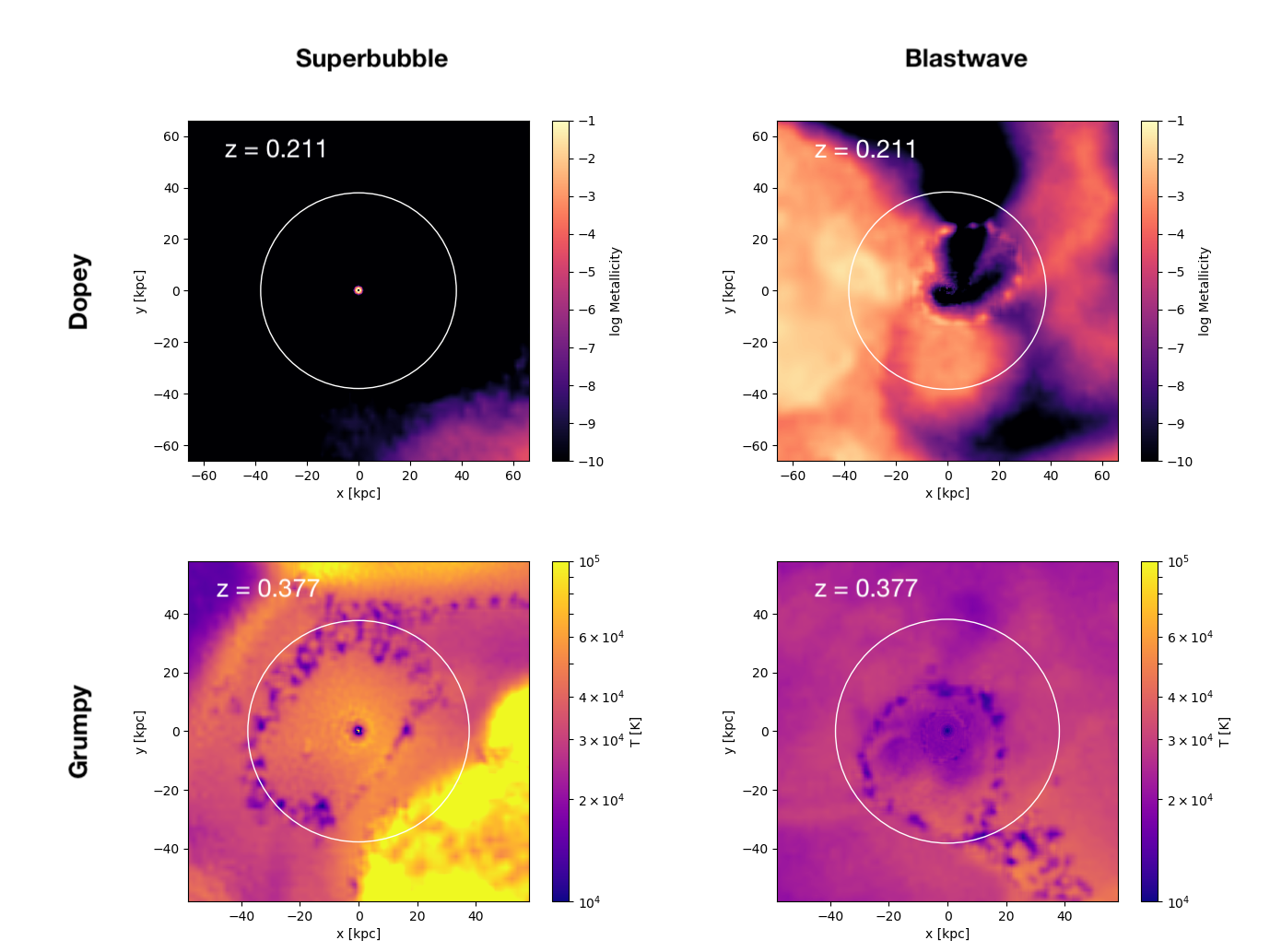}
\caption{
Upper row: metallicity slice of gas around Dopey's dark halo in our simulation with the superbubble feedback model (left) and in the simulations of \shenonex, with the blastwave feedback model (right), at redshift $z=0.211$. 
Bottom row: temperature slice of gas within Grumpy's dark halo in our simulation with the superbubble feedback model (left) and in the simulations of \shenonex, with the blastwave feedback model (right), at redshift $z=0.377$. 
The virial radii of Dopey and Grumpy are marked by the white circles.
}\label{Fig:DopeyGrumpy}
\end{figure*}

Dopey and Grumpy, the two faint dwarf galaxies found in \citet{2014ApJ...792...99S}, do not form in our simulation. The gas in their halos appear not cold and dense enough to ignite star formation. Although the dark matter halo assembly for Dopey and Grumpy in our simulation are nearly identical to those in \citet{2014ApJ...792...99S}, the halos accrete less gas, as shown in Fig. \ref{Fig:baryonfrac_evo} (the purple and green lines for Dopey and Grumpy, respectively). The reduction of gas accretion occurs early (around 2-3 Gyr), and lasts throughout cosmic history. As Dopey and Grumpy did not form stars themselves, this reduction is entirely due to feedback from the nearby more massive galaxies, Bashful and Doc. As shown in \citet{2020MNRAS.493.2149K}, the entropy-driven outflows unique to the superbubble feedback generally have higher temperature and higher mass loading at the ISM. And because Dopey and Grumpy are only 100 - 250 kpc away from the Bashful and Doc at majority of the time(well within the influence of galactic winds from the two more massive galaxies), the outflows heat the CGM of these small halos to temperatures of order the virial value, suppressing their gas accretion, and subsequently lower halo gas densities, producing longer cooling time (as $t_{\rm cool} \propto 1/\rho$). In other words, galactic winds from Bashful and Doc serve as a form of ``preventive feedback'' for Dopey and Grumpy. 

In addition, Dopey and Grumpy are influenced by Bashful and Doc in other ways, which further prevent gas from condensing into the centres of their potential wells. In Fig.~\ref{Fig:DopeyGrumpy} we compare the gas properties within the dark halos where Dopey (upper row) and Grumpy (bottom row) are supposed to form, prior to their formation times, in case the stellar feedback is described by the superbubble (left column) and the blastwave (right column) models.

\subsection{Dopey - absence of CGM metal transfer}
In \citet{2014ApJ...792...99S}, Dopey starts forming the first stars around redshift $z \sim 0.211$, and the star formation is triggered by an interaction with a nearby dark halo. The gas accreted by its dark matter halo has metallicity of order $10^{-2}-10^{-1} Z_{\odot}$ (as shown in the upper right panel of Fig.~\ref{Fig:DopeyGrumpy}). 
This is drastically different in our simulation. As shown in the upper left panel of Fig.~\ref{Fig:DopeyGrumpy}, the halo gas of Dopey is largely primordial at $z=0.211$, which further reduce the cooling rate compared to the blastwave simulation, preventing gas from forming stars.  
Since Dopey has not formed any stars yet in either case, the metal enrichment of Dopey's halo gas are from the CGM of Doc and Bashful. This ``intergalactic metal transfer'' is also seen in cosmological simulations of more massive systems \citep{Shen:2011cs,2017MNRAS.468.4170M,10.1093/mnras/stx1517}, but as galactic winds from dwarf galaxies usually have higher mass and metal loading factor \citep[e.g.,][]{Christensen_2018}, majority of metals produced in these galaxies are ejected to the CGM, thus metal transfer between CGM is more likely occur in field dwarf systems. The absence of enriched gas in Dopey's halo in our simulation implies that the extent of the metal-enriched CGM is smaller with the superbubble feedback than the one with the blastwave model. This is again due to the difference in feedback models. We will explore the CGM in more detail in Section~\ref{sec:cgm}.

\subsection{Grumpy - here it was the butler}
For the case of Grumpy, the situation is somewhat different than for Dopey. By the time Grumpy formed in the previous simulation with the blastwave feedback (at $z \sim 0.377$), 
the halo gas in Grumpy has higher average temperature ($T \sim 4.3\times 10^4$ K) than the blastwave case ($T \sim 2.1\times 10^4$ K) in \citet{2014ApJ...792...99S}. As shown in the lower panels of Fig. \ref{Fig:DopeyGrumpy}, some gas reaches $\sim 10^{5}$ K right outside the virial radius. The average halo gas density is also significantly lower, with number density $n \sim 2.4 \times 10^{-5}$ atoms/cm$^3$ instead of $9.4 \times 10^{-5}$ atoms/cm$^3$ in the blastwave run.

In addition to the prevention of gas accretion discussed before, here the heating and dispersion of Grumpy's halo gas is a direct result of the active star formation in Bashful at low redshift in our simulation. Multiple galactic winds driven by supernova explosions in Bashful have swept through Grumpy's halo right before $z=0.377$, one is clearly visible in Fig. \ref{Fig:DopeyGrumpy} and an earlier wind was found around $z \sim 0.67$, after a strong starburst in Bashful. The combination of heating and decreased density prevents gas from cooling and condensing into the halo center and forming stars. Since the gas remains diffuse, strong ram-pressure stripping occurs when Grumpy interacts with
Doc (as shown in the baryon fraction ``dip'' around 12.5 Gyr in Fig.~\ref{Fig:baryonfrac_evo}). At later times, gas re-accretes, and at $z=0$ the halo has total gas mass of $M_{\rm gas} = 9.3 \times 10^{6} M_{\odot}$ a baryon fraction of 0.005, not very different from Dopey. However the gas remains diffuse and warm, unable to ignite star formation. 

In \citet{2014ApJ...792...99S}, Dopey and Grumpy lack an old stellar population, have blue colors and very compact morphologies, similar to the properties of extremely metal-deficient compact dwarfs (XBCDs) \citep{2008A&A...491..113P}. The failure of the formation of Dopey and Grumpy in our simulation underlines the fact that the formation such low mass dwarf galaxies are highly stochastic, and likely to be sensitive to physical processes such as gravitational interactions, feedback, ionising radiation from UV background and local sources, and star formation models \citep[e.g.,][]{Munshi_2019}. However, it is important to mention that our simulation describes a particular case and not all galaxies hosted by halos with $M_{\rm vir} < 10^{10}~M_{\odot}$ are influenced by neighbouring galaxies like in our groups of dwarfs. It is also worth noting that the lack of pre-reionisation star formation can be an artefact from resolution and how reionisation is modelled in our simulation (UV background is turned on instantaneously). Although in \citet{2014ApJ...792...99S}, star formation did not occur even without UV beackground in a twin simulation by the time reionisation is complete ($z \sim 6$), higher resolution may enable gas self-shielding more realistically which allows gas to have runaway cooling and condensation. Some recent works \citep[e.g.,][Applebaum et al.  in prep.]{10.1093/mnras/stv2072, 2017MNRAS.471.3547F, Wheeler2019,2019MNRAS.487.3740W} do form ultra-faint dwarf galaxies with predominately ancient stellar population. We reserve for the future detailed investigations of the effects of superbubble feedback, non-uniform UV radiation and self-shielding using higher resolution simulations of a larger population of dwarf galaxies.

\section{The Circumgalactic Medium}
\label{sec:cgm}
\begin{figure*}
\includegraphics[width=\linewidth]{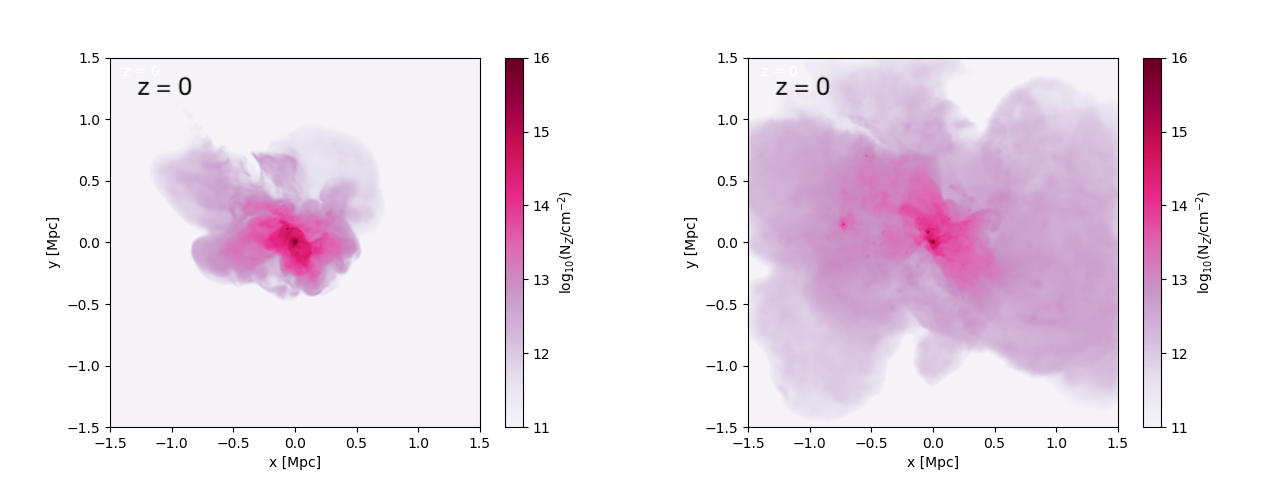}
\caption{
Present day column densities for metals computed, in a region of $3$ Mpc on a side, for our simulation with the superbubble feedback model (left panel), and for the one computed in \shenonex, with the blastwave feedback model (right panel). 
}\label{Fig:col_dens_Z}
\end{figure*}

\begin{figure*}
\includegraphics[width=\linewidth]{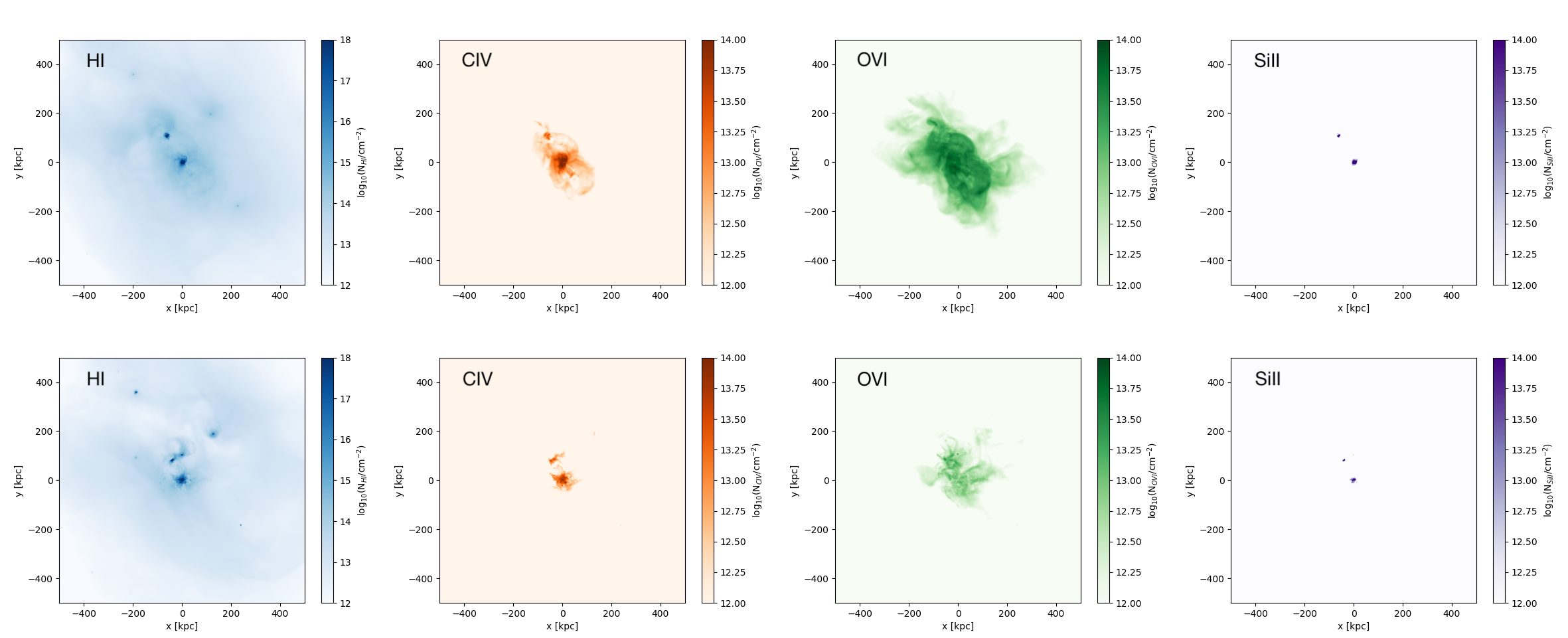}
\caption{
Projections of the present day column densities computed, in a region of $1$ Mpc on a side, for ions of different species: \ion{H}{i}, \ion{C}{iv}, \ion{O}{vi} and \ion{Si}{ii}. The upper row shows the present day column densities computed for our simulation, with the superbubble feedback model. Instead, the bottom row shows the ones computed in \shenonex, with the blastwave feedback model.
}\label{Fig:col_dens}
\end{figure*}

\begin{figure*}
\includegraphics[width=\linewidth]{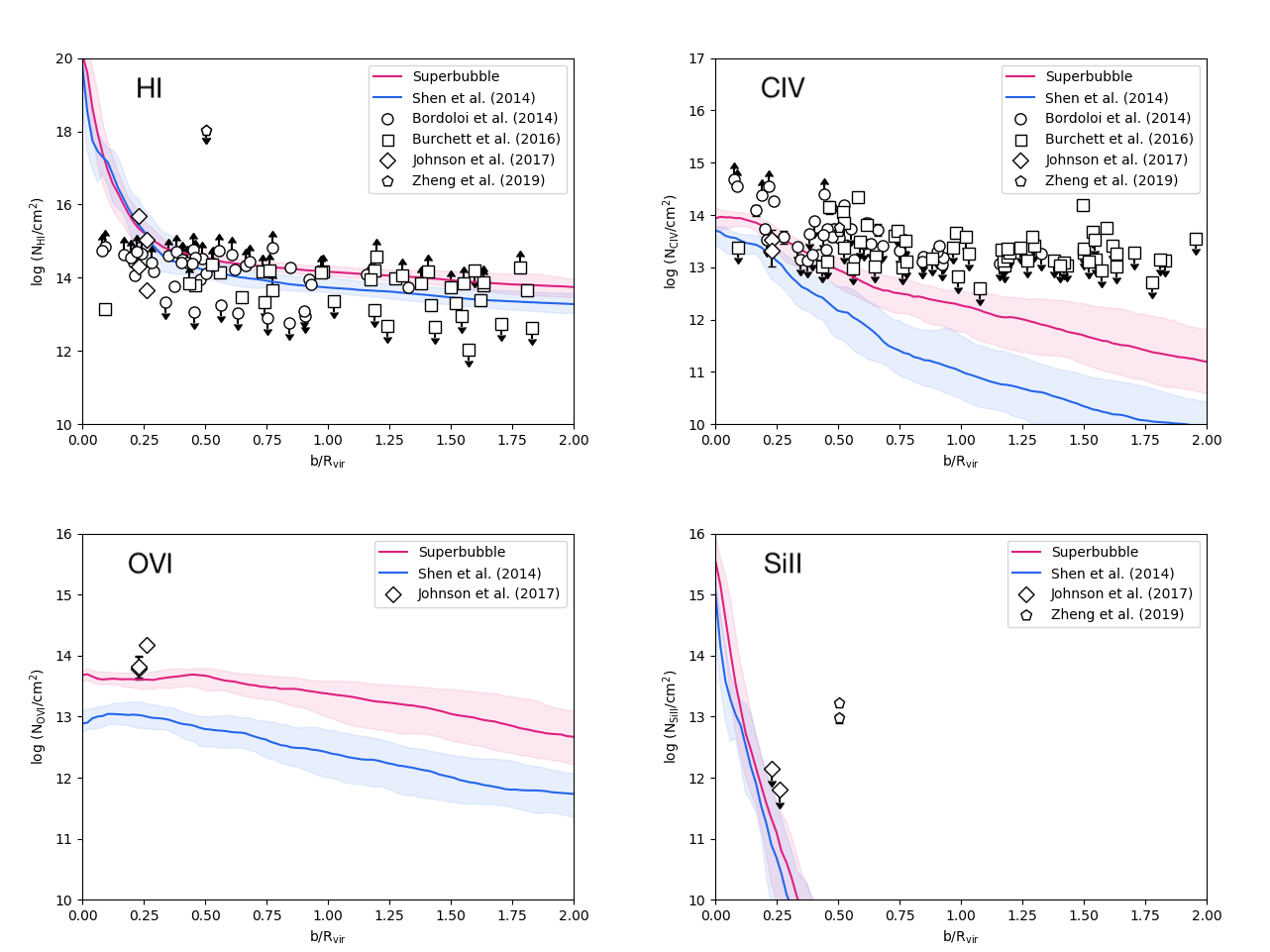}
\caption{
Present day column densities for ions of different species: \ion{H}{i} (upper left), \ion{C}{iv} (upper right), \ion{O}{vi} (bottom left) and \ion{Si}{ii} (bottom right). On the $x$-axis, the impact parameter $b$ is normalised by the virial radius of Bashful. 
The solid lines represent the present day column densities computed for our simulation (red), with the superbubble feedback model, and for the simulation presented in \shenone (blue), with the blastwave feedback model. The empty symbols correspond observations of the CGM around COS-Halos and COS-Dwarfs galaxies \citep[empty circles][]{2014ApJ...796..136B}, nearby faint dwarf galaxies \citep[empty squares][]{2016ApJ...832..124B}, star-forming field dwarf galaxies \citep[empty diamonds][]{,2017ApJ...850L..10J}, and the low-mass dwarf galaxy WLM \citep[empty pentagon][]{2019MNRAS.490..467Z}.
}\label{Fig:column_dens}
\end{figure*}

Because of their shallower potential wells, the formation and evolution of dwarf galaxies are highly dependent to how these systems and their surrounding intergalactic medium (IGM) exchange energy, mass and metals. For example, as shown in Section \ref{sec:mzr}, strong galactic outflows eject $68.8\%$ and $63.3\%$ of the metals ever produced in Bashful and Doc, respectively, into the CGM and the IGM. The bursty star formation history (Fig.~\ref{Fig:SFH}) and the rapid changes of baryonic mass at the inner part of the galaxies (Fig.~\ref{fig:Mb500_time}) show that galactic outflows can temporary evacuate the ISM. The mass loading factor of galactic winds at the virial radii, defined as the mass ejection rate per unit star formation rate ($\dot{M}_{\rm w}/\dot{M}_{*}$) often reach to values $>10$, much higher than that for more massive galaxies \citep[e.g.,][]{Shen_2013,Christensen_2018}. In turn, the properties of these outflows determine subsequent gas accretion and star formation, and the next cycle of baryonic flows. In this section, we briefly investigate the chemical imprints of the baryon cycles in the multi-phase CGM, and compare our simulation with recent observations from absorption line studies of near field dwarf galaxies. As we show in the following paragraph, the CGM remains a sensitive probe to feedback mechanisms: the ionisation and chemical properties of the CGM in our simulation differs significantly from those in \citet{2014ApJ...792...99S} with the blastwave feedback model. 

Fig.~\ref{Fig:col_dens_Z} shows the total column density ($N_{\rm Z}$) of metals at $z=0$ for our simulation (left panel) and the result from \citet{2014ApJ...792...99S} with the blastwave feedback (right panel). The extent of the metal-enriched CGM, defined as the impact parameter from the center of Bashful at which the covering fraction of $N_{\rm Z} > 10^{12}$ cm$^{-2}$ drops below $50$\%, is about $481$ kpc, approximately 4.7 times Bashful's virial radius ($R_{\rm vir, B}$) at $z=0$. Galactic outflows are clearly very effective enriching the CGM. However, the size of the metal bubble is much smaller than that from the blastwave model (which reaches $16 R_{\rm vir, B}$ at $z=0$), even though the total mass of metals ejected are similar for both simulations. In many ways, the superbubble feedback is less ``ejective'' for the halo gas, and this is also reflected in the higher baryonic fractions and higher $M_{\rm HI}/M_{*}$ (for Bashful and Doc, as shown in Fig.~\ref{Fig:gas_lum} and Fig.~\ref{Fig:baryonfrac_evo}). This is also consistent with the theoretical model proposed by \citet{2020MNRAS.493.2149K}, where the entropy-driven winds have relatively lower velocities. 

In addition to the differences in feedback, the extent of metal bubble is likely also a result from the star formation histories of the two galaxies, especially Bashful. 
In our simulation, Bashful forms about 50\% of its total stellar mass by $z \gtrsim 1$, whereas the blastwave counterpart forms more than 80\% in the same period.
Earlier star formation and outflows are more likely to escape the halo potential well and reach larger distances.     

In Fig.~\ref{Fig:col_dens}, we compare the column density maps of ionisation species \ion{H}{i}, \ion{C}{iv}, \ion{O}{vi}, and \ion{Si}{ii} from our simulation as well as from the simulation in \citet{2014ApJ...792...99S}. We compute the ionisation fraction of each element using \textsc{Cloudy} \citep{1998PASP..110..761F}, assuming the gas is under a uniform UV beackground at $z=0$ from \citet{2012ApJ...746..125H}. As the star formation rates at $z=0$ for both galaxies are of order $10^{-2}$ M$_{\odot}$/yr, we do not expect internal radiation to have a significant impact the ionisation state of CGM. The column density is computed using the same analysis tool as described in \citet{Shen_2013}. In Fig. \ref{Fig:column_dens} we compare the 1D column density distribution of these ions as a function of the impact parameter normalised by the galaxies' virial radii ($b/R_{\rm vir}$) with observed values from recent surveys targeted on low mass dwarf galaxies \citep{2014ApJ...796..136B,2016ApJ...832..124B,2017ApJ...850L..10J,2019MNRAS.490..467Z}. 

The column densities for different ions exhibit very different distributions. First, we note that neutral hydrogen (\ion{H}{i}) is pervasive. At $b/R_{\rm vir} \sim 1$, the median column density of \ion{H}{i} around our dwarf galaxies is N(\ion{H}{i}) $= 1.5\times 10^{14} $cm$^{-2}$, and the covering fraction for absorption systems with N(\ion{H}{i}) $> 10^{13.5} $cm$^{-2}$ within $R_{\rm vir}$ is $100$\%. This is consistent with observations, which often find that \ion{H}{i} Ly$\alpha$ absorption is ubiquitous around galaxies with a large range of stellar mass \citep[e.g., ][]{2014ApJ...792....8W,2016ApJ...832..124B, Bordoloi_2018}. For example, \citet{2016ApJ...832..124B} find a very high detection rate (16 out of 21) of systems with N(\ion{H}{i}) $> 10^{13.5} $cm$^{-2}$ from the COS-Dwarfs survey for galaxies of stellar mass range log$M_{*}/M_{\odot} = 8-9$. The 1D column density distribution for \ion{H}{i} from our simulation also shows an excellent agreement with the COS-Dwarfs data from \citet{2014ApJ...796..136B} and \citet{2016ApJ...832..124B}. It is interesting to note that the N(\ion{H}{i})-b relationship from the \citet{2014ApJ...792...99S} simulation using the blastwave feedback model is very similar to the present work, which indicate that, unlike metal ions, the distribution of \ion{H}{i} is less sensitive to feedback mechanisms (provide that the feedback launches sufficient outflows). The blastwave model does produce a lower \ion{H}{i} column density in the CGM, consistent with the picture that gas (and metals) are spread to a larger distances indicated by the metal bubble comparison from Fig. \ref{Fig:col_dens_Z}. 

In contrast to \ion{H}{i}, the enriched cold gas appears to be concentrated within the galaxies. For example, the column densities of \ion{Si}{ii} (tracing cold, T $\sim 10^{4}$K gas) decrease rapidly to below $10^{12}$ cm$^{-2}$ within $15.9$ kpc from Bashful's center (i.e., $0.19 \ R_{\rm vir,B}$). This is because the diffuse CGM of low mass galaxies like Bashful and Doc is highly ionised by the UV background, and majority of the metals are in fully ionised form.
The silicon ions tracing the cool diffuse gas, \ion{Si}{ii}, \ion{Si}{iii} and \ion{Si}{iv}, only account for $27.9$\% of the total silicon budget in the CGM. 
This is also in agreement with recent CGM observations on field dwarf galaxies of similar mass from \citet{2017ApJ...850L..10J}, in which they found non-detection of \ion{Si}{ii} (N(\ion{Si}{ii})$\la 10^{12}$ cm$^{-2}$) around two galaxies (D1 and D2 in their paper) with stellar mass around $10^{8} M_{\odot}$. The low and intermediate silicon ions (\ion{Si}{2-4}) add up to 2\%-6\% of the total silicon mass. \citet{2019MNRAS.490..467Z}, on the other hand, did detect \ion{Si}{ii} at a large impact parameter (45.2 kpc) around WLM (which has $M_{*}=4.3 \times 10^{7} M_{\odot}$), with column density around $10^{13} \  \rm cm^{-2}$. A larger sample of observations and simulations on the CGM around faint field dwarf galaxies is probably needed to reach statistically significant conclusions.   

Column density distribution of higher ionisation species, such as \ion{C}{iv} or \ion{O}{vi}, appear to be more extended than \ion{Si}{ii}. For the moderately ionised \ion{C}{iv}, the median column density drops below $10^{13.5} \ \rm cm^{-2}$ (a common detection threshold) at $0.3 R_{\rm vir, B}$, about $24.75$ kpc. At larger distances it appears mostly below the observational points. However, observations from \citet{2016ApJ...832..124B} shows that \ion{C}{iv} detection is associate with galaxies of stellar mass $> 10^{9.5} M_{\odot}$, and is very rare for lower mass galaxies, and this is consistent with what we find here. Our results are also in agreement with the D1 and D2 system in \citet{2017ApJ...850L..10J}, although again below the WLM data point. The column density of highly ionized \ion{O}{vi} decreases slower with impact parameter, and remains detectable (N(\ion{O}{vi})$ > 10^{13.5} \ \rm cm^{-2}$) up to $\sim R_{\rm vir}$. Still, the mass in \ion{O}{vi} accounts for merely $11.5$\% of the total oxygen mass in the CGM. 
The column densities from our simulation compare reasonably well with those in \citet{2017ApJ...850L..10J}, albeit slightly lower. For both \ion{C}{iv} and \ion{O}{vi}, the column densities from the simulation in \citet{2014ApJ...792...99S} with the blastwave feedback are significantly lower (1 dex or more at larger impact parameters) than those our simulation. Since the diffuse CGM temperature is generally around a few times $10^{4}$ K, the main channel of producing intermediate and high ions is photo-ionisation from the UV beackground. 
Thus, although the global quantities such as stellar mass, size, kinematics, total metallicity remains similar for runs with different feedbacks, the CGM proved to be a more sensitive probe of feedback mechanisms.  

In general, although dwarf galaxies eject majority of the metals into the CGM and the IGM, the column densities of commonly observed metal ions are lower than those associated with more massive galaxies at similar redshift, especially for lower ionisation species \citep[e.g.,][]{10.1093/mnras/sts702,10.1093/mnras/stw1066,10.1093/mnras/sty656}, and this trend is confirmed by observation data. This is likely due to several factors. First, field dwarf galaxies reside in a low density environment, have lower star formation rate and produce a smaller amount of metals. These metals are spread over a larger volume, so the metallicity of the CGM gas is lower. Second, both accretion flows and outflows appear to be diffuse. At least in our simulation, there is neither clear filamentary accretion nor clumpy outflows, both of which would increase the column density and covering fraction of low ionisation lines \citet[e.g.,][]{Shen_2013}. Third, the low density environment makes it easier for the gas to be photoionized to higher states, thus the common lines only traces a small fraction of the total metal budget. For ions like \ion{O}{vi}, the origins around dwarf galaxies may differs significantly from those around $L_{*}$ galaxies, with the latter being dominated by gas heated from feedback or accretion shocks cooling through the window of $T\sim10^{5.5}$ K (i.e. peak of \ion{O}{vi} abundance in collisionally ionisation models) \citep[][Spilker er al. in preparation]{2019MNRAS.484.3625R}. We defer a more quantitative study of CGM across different mass galaxies in a forthcoming paper.

\section{Conclusions}\label{sec:concl}
In this work, we presented the results of a hydrodynamic ``zoom-in'' simulation of dwarf galaxies within the $\Lambda$CDM model, performed with the code \codenamex. 
The simulation includes a star formation recipe based on the Schmidt law, with high gas density threshold, a metal-dependent radiative cooling and a uniform UV background, which modifies the ionization and excitation state of the gas, and a scheme for the turbulent diffusion of metals and thermal energy.
The stellar feedback follows the novel superbubble model, which includes a sub-resolution treatment of multi-phase gas particles, thermal conduction between cold and hot phases, and a model for stochastic evaporation of cold clouds \citep{2014MNRAS.442.3013K}. 
The simulation adopts the same initial conditions and numerical setup as in \shenonex, with the prescription for the supernova feedback being the only difference. 

Within our simulation volume, only two luminous dwarfs have formed by redshift $z=0$, namely Bashful and Doc. If they were to be observed today, Bashful and Doc would appear blue and classified as dIrrs.
We compared the two simulated dwarfs with the results of \shenone and \shentwox, in order to qualitatively study the difference between the superbubble and the blastwave feedback models.
We also compared them with dwarf galaxies observed in the Local Volume and with the results presented in other recent numerical investigations of galaxy formation and evolution.
In particular, we analysed the stellar and cold gas properties of the two luminous dwarf galaxies, their star formation histories, their cusp-core transformation, and the metal enrichment of their surrounding CGM. In addition, we explored why the two faint galaxies formed in \citet{2014ApJ...792...99S} failed to form in less massive halos ($M_{\rm vir} < 10^{10} \rm M_{\odot}$) in our simulation.

We summarise the main findings of our work in the following: 
\begin{itemize}
    \item In general, Bashful and Doc have present day realistic properties. They have low star formation efficiencies, high cold gas fractions and low metallicities, closely following the empirical scaling relations found in dwarf galaxies observed in the Local Volume. The superbubble feedback  efficiently suppresses star formation in lower mass galaxies, resulting a factor of 2 less stellar mass in Doc comparing to the delay-cooling model. 
    
    \item Both Bashful and Doc have highly bursty star formation histories due to the temporary evacuation of the ISM by galactic winds. Although the SFHs are qualitatively similar to the ones of the same dwarfs in \shenonex, the entropy-driven winds from the superbubble feedback are less ejective (at the halo/CGM scale), resulting in  larger baryonic fraction and a larger $M_{\rm HI}/M_{*}$ ratio for both galaxies. The superbubble feedback also generates less turbulence in the HI disk, leading to a higher $v_{\rm rot}/\sigma$ ratio.      
    
    \item The rapid cycles of depletion and re-accretion of gas in the innermost regions of the halos lead to strong fluctuations of the central gravitational potential, causing the central dark matter cusps to flatten into cored density profiles. At z = 0, the DM core sizes for Bashful and Doc are 3.45 and 1.86 kpc, respectively. The superbubble model appears more effective in destroying dark matter cusps than the previously adopted delayed cooling model. For the same amount of stellar mass formed, the superbubble feedback can remove up to 2-3 times more mass from the cusp, suggesting a higher coupling efficiency between the energy injected by SNe and the ISM. Likewise, Bashful and Doc also have larger stellar disks in our simulation than in \shenonex. The strong correlation between the DM core size and the disk scale lengths is consistent with previous work by \citet{2012ApJ...759L..42P} and also with observations \citep{2009Natur.461..627G}. 

    \item In \shenonex, two additional UFD galaxies form, namely Dopey and Grumpy. In our simulation, the two corresponding dark halos never accrete gas cold and dense enough to ignite star formation, and they remain dark due to the influence of Bashful and Doc. The superbubble-driven galactic winds from Bashful and Doc are hotter and more pressurized, which suppress gas accretion to the nearby halos for Dopey and Grumpy. In addition, cooling time is further increased for Dopey due to an absence of CGM metal transfer (existed in \shenonex). The halo gas of Grumpy is first swept by winds from Bashful, then ram-pressure stripped due to the close interactions with Doc, again leading to a longer cooling time and failure of gas condensation. Our results suggest that the formation of UFD galaxies strongly depends upon the stellar feedback and environmental effects.
    
    \item Due to the shallow potential wells of dwarf galaxies, galactic outflows are very effective in enriching the CGM. In our simulation, the metal-enriched CGM extends to about 4.7R$_{\rm vir}$ of Bashful, significantly smaller than the one in \shenonex. This again suggests the superbubble feedback is less ejective at the CGM scale (although very effective ejecting gas at the galaxy center, as shown in Figure \ref{fig:Mb500_time}), consistent with the analytical model of \citet{2020MNRAS.493.2149K}. The smaller extent of metal ``bubble'' directly results in the aforementioned absence of metals in Dopey's halo.  
    
    \item The column density distributions of \ion{H}{i}, \ion{Si}{ii}, \ion{C}{iv} and \ion{O}{vi} as a function of scaled impact parameter ($b/R_{\rm vir}$) are in good agreement with recent observations of CGM around isolated dwarf galaxies. While \ion{H}{i} is ubiquitous with a covering fraction of unity within the CGM, low and intermediate ions like \ion{Si}{ii} and \ion{C}{iv} are less extended (typically confined within $0.2-0.3~R_{\rm vir}$), and non-detections are common. \ion{O}{vi} is more extended with column density N(\ion{O}{vi})$ \ga 10^{13.5}$ cm$^{-2}$ within $R_{\rm vir}$, but the mass in \ion{O}{vi} is only 11\% of the total CGM oxygen budget, as the diffuse CGM is highly ionised by the UV background. Superbubble feedback produces \ion{C}{iv} and \ion{O}{vi} column densities that are an order of magnitude higher than those with blastwave feedback. 
    
\end{itemize}

In summary, the superbubble model better describes the physics behind the supernova feedback process and it is capable of forming dwarf galaxies which closely resemble real ones, reproducing the stellar mass and the cold gas content, bursty star formation histories, the stellar kinematics, and metallicities of dwarf galaxies observed in the Local Volume. 
The impact of SN feedback on the host dark matter halos and the metal enrichment of the CGM represent the most sensitive probes of feedback mechanisms. 

Future work will aim to improve the results obtained here, perhaps increasing the simulation volume to have a larger sample of dwarf galaxies.

\begin{acknowledgements}
The analysis was partially done with \texttt{PyNBODY} \citep{2013ascl.soft05002P}, and we thank Andrew Pontzen for providing this very useful analysis tool. We thank Bernhard Baumschlager, Claudia Cicone, Davide Decataldo and Robert Wissing for very helpful and interesting discussions. The simulations were performed using the resources from the DiRAC HPC facilities in the UK, and the National Infrastructure for HPC and Data Storage in Norway, UNINETT Sigma2 (Project NN9477K). S. Shen acknowledges the support from the Research Council of Norway through NFR Young Research Talents Grant 276043.  
BW Keller acknowledges funding in the form of a Postodctoral Research Fellowship from the Alexander von Humboldt Siftung. PM acknowledges  a NASA contract supporting the WFIRST-EXPO Science Investigation Team (15-WFIRST15-0004), administered by GSFC. 
\end{acknowledgements}

\bibliography{references}

\end{document}